\documentclass[preprint2]{aastex}

\shorttitle{Proto-groups in zCOSMOS-deep}
\shortauthors{Diener et al.}

\begin{document}

\title{Proto-groups at $1.8<z<3$ in the zCOSMOS-deep sample}
\author{C. Diener\altaffilmark{1},
S.J. Lilly\altaffilmark{1}, 
C. Knobel\altaffilmark{1}, 
G. Zamorani\altaffilmark{2},
G. Lemson\altaffilmark{18},
P. Kampczyk\altaffilmark{1},
N. Scoville\altaffilmark{3}, 
C. M. Carollo\altaffilmark{1},
T. Contini\altaffilmark{4,5},
J.-P. Kneib\altaffilmark{6},
O. Le Fevre\altaffilmark{6},
V. Mainieri\altaffilmark{7},
A. Renzini\altaffilmark{8},
M. Scodeggio\altaffilmark{9},
%
S. Bardelli\altaffilmark{2},
M. Bolzonella\altaffilmark{2},
A. Bongiorno\altaffilmark{10},
K. Caputi\altaffilmark{1,21},
O. Cucciati\altaffilmark{11},
S. de la Torre\altaffilmark{12},
L. de Ravel\altaffilmark{12},
P. Franzetti\altaffilmark{9},
B. Garilli\altaffilmark{9},
A. Iovino\altaffilmark{11},
K. Kova\v{c}\altaffilmark{1},
F. Lamareille\altaffilmark{4,5},
J.-F. Le Borgne\altaffilmark{4,5},
V. Le Brun\altaffilmark{6},
C. Maier\altaffilmark{1,20},
M. Mignoli\altaffilmark{2},
R. Pello\altaffilmark{4,5},
Y. Peng\altaffilmark{1},
E. Perez Montero\altaffilmark{4,5,13},
V. Presotto\altaffilmark{11},
J. Silverman\altaffilmark{14},
M. Tanaka\altaffilmark{14},
L. Tasca\altaffilmark{6},
L. Tresse\altaffilmark{6},
D. Vergani\altaffilmark{2,22},
E. Zucca\altaffilmark{2},
%
%
R. Bordoloi\altaffilmark{1},
A. Cappi\altaffilmark{2},
A. Cimatti\altaffilmark{15},
G. Coppa\altaffilmark{10},
A. M. Koekemoer\altaffilmark{16},
C. L\'opez-Sanjuan\altaffilmark{6},
H. J. McCracken\altaffilmark{17},
M. Moresco\altaffilmark{15},
P. Nair\altaffilmark{2},
L. Pozzetti\altaffilmark{2},
and N. Welikala\altaffilmark{19}
}

\altaffiltext{1}{Institute for Astronomy, Department of Physics, ETH Zurich, Zurich 8093 Switzerland}
\altaffiltext{2}{INAF Osservatorio Astronomico di Bologna, via Ranzani 1, I-40127, Bologna, Italy}
\altaffiltext{3}{California Institute of Technology, MC 105-24, 1200 East California Boulevard, Pasadena, CA 91125}
\altaffiltext{4}{Institut de Recherche en Astrophysique et Plan\'etologie, CNRS, 14, avenue Edouard Belin, F-31400 Toulouse, France}
\altaffiltext{5}{IRAP, Universit\'e de Toulouse, UPS-OMP, Toulouse, France}
\altaffiltext{6}{Laboratoire d'Astrophysique de Marseille, CNRS/Aix-Marseille Universit\'e, 38 rue Fr\'ed\'eric Joliot-Curie, 13388, Marseille cedex 13, France}
\altaffiltext{7}{European Southern Observatory, Garching, Germany}
\altaffiltext{8}{INAF-Osservatorio Astronomico di Padova, Vicolo dell'Osservatorio 5, 35122, Padova, Italy}
\altaffiltext{9}{INAF-IASF Milano, Milano, Italy}
\altaffiltext{10}{Max Planck Institut f\"ur Extraterrestrische Physik, Garching, Germany}
\altaffiltext{11}{INAF Osservatorio Astronomico di Brera, Milan, Italy}
\altaffiltext{12}{Institute for Astronomy, University of Edinburgh, Royal Observatory, Edinburgh, EH93HJ, UK}
\altaffiltext{13}{Instituto de Astrofisica de Andalucia, CSIC, Apartado de correos 3004, 18080 Granada, Spain}
\altaffiltext{14}{Institute for the Physics and Mathematics of the Universe (IPMU), University of Tokyo, Kashiwanoha 5-1-5, Kashiwa, Chiba 277-8568, Japan}
\altaffiltext{15}{Dipartimento di Astronomia, Universit\`a degli Studi di Bologna, Bologna, Italy}
\altaffiltext{16}{Space Telescope Science Institute, Baltimore, MD 21218, USA}
\altaffiltext{17}{Institut d'Astrophysique de Paris, UMR7095 CNRS, Universit\'e Pierre \& Marie Curie, 75014 Paris, France}
\altaffiltext{18}{Max Planck Institut f\"ur Astrophysik, Garching, Germany}
\altaffiltext{19}{Institut d'Astrophysique Spatiale, B\^atiment 121, Universit\'e Paris-Sud XI and CNRS, 91405 Orsay Cedex, France}
\altaffiltext{20}{University of Vienna, Department of Astronomy, Tuerkenschanzstrasse 17, 1180 Vienna, Austria}
\altaffiltext{21}{Kapteyn Astronomical Institute, University of Groningen, P.O. Box 800, 9700 AV Groningen, The Netherlands}
\altaffiltext{22}{INAF-IASF Bologna, Via P. Gobetti 101, I-40129 Bologna, Italy
}

\begin{abstract}
We identify 42 ``candidate groups'' lying between $1.8 < z < 3.0$ from a sample of 3502 galaxies with spectroscopic redshifts in the zCOSMOS-deep redshift survey within this same redshift interval. These systems contain three to five spectroscopic galaxies that lie within 500\,kpc in projected distance (in physical space) and within 700\,km/s in velocity. Based on extensive analysis of mock catalogues that have been generated from the Millennium simulation, we examine the likely nature of these systems at the time of observation, and what they will evolve into down to the present epoch.  Although few of the ``member'' galaxies are likely to reside in the same halo at the epoch we observe them, 50\% of the systems will have, by the present epoch, all of the member galaxies in the same halo, and almost all (93\%) will have at least some of the potential members in the same halo.  Most of the candidate groups can therefore be described as ``proto-groups''.  A crude estimate of the overdensities of these structures is also consistent with the idea that these systems are being seen as they assemble.  We also examine present-day haloes and ask whether their progenitors would have been seen amongst our candidate groups. For present-day haloes between $10^{14}-10^{15}$\,M$_{\sun}$/h, 35\% should have appeared amongst our candidate groups, and this would have risen to 70\% if our survey had been fully-sampled, so we can conclude that our sample can be taken as representative of a large fraction of such systems.  There is a clear excess of massive galaxies above $10^{10}$\,M$_{\sun}$ around the locations of the candidate groups in a large independent COSMOS photo-$z$ sample, but we see no evidence in this latter data for any color differentiation with respect to the field.  This is however consistent with the idea that such differentiation arises in satellite galaxies, as indicated at $z < 1$, if the candidate groups are indeed only starting to be assembled.

\end{abstract}


\section{Introduction}
Groups of galaxies, by which we mean sets of galaxies that occupy the same dark matter halo, are important for several reasons. They constitute the largest virialized systems in the universe and are therefore probes for the growth of structure and eventually the underlying cosmological model.
Furthermore, groups provide an environment different from the field. The group environment is suspected of influencing the evolution and properties of the member galaxies through various processes as ram pressure stripping (Gunn \& Gott 1972, Dressler 1980, Abadi et al. 1999), strangulation (Larson et al. 1980, Kawata \& Mulchaey 2008), enhanced merger rate (Spitzer \& Baade 1951), galaxy harassment (Moore et al. 1996) and so on. Recent work at low redshift (Peng et al. 2010, 2012, Prescott et al. 2011, Weinmann et al. 2009, van den Bosch et al. 2008) has indicated that the dominant process producing environmental differentiation in the galaxy population at low redshift (at least as regards the fraction of galaxies in which star-formation has been ``quenched'') is arising from changes to satellite galaxies and there is evidence that this is true also out to $z\sim1$ (Knobel et al. 2012, Kovac et al. 2012).  Various papers have established the influence of the group environment on the galaxy population by investigating the morphology-density relation (Oemler 1974, Balogh et al. 2004) or the differences between centrals and satellites (Peng et al. 2012, Pasquali et al. 2010, Skibba 2009).

Identifying groups using discrete galaxies as a tracer sample is a non-trivial task. Previous work at low and intermediate redshift discusses extensively the performance of different group finders, in terms of the underlying dark matter haloes. Common automated group finding methods are the friends-of-friends method (Huchra \& Geller 1982, Eke et al. 2004, Berlind et al. 2006), the Voronoy-Delaunay method (Marinoni et al. 2002, Gerke et al. 2005, Cucciati et al. 2010) or a combination of both (Knobel et al. 2009, 2012).

Little is known about groups at $z > 1$, mostly because few redshift surveys have penetrated beyond this depth with a high enough sampling density to have any hope of finding any except the most massive.  The redshift interval around $z \sim 2$ is of interest for several reasons. This is, as will be clear in this paper, when the first groups consisting of multiple massive (around M*) galaxies should appear in the Universe in significant numbers.  It is also close to the peak of star-formation (Hopkins \& Beacom 2006, Reddy et al. 2008) and AGN activity (Wolf et al. 2003) in the Universe, and where we might expect the first effects of the environment in controlling galaxy evolution to become apparent.

Above a redshift of $z\sim2$ there exist only rare examples of single clusters or groups in the literature. The search for them relies on overdensities around radio galaxies (Miley et al. 2006, Venemans et al. 2007), the search for X-ray emission (Gobat et al. 2011) as well as overdensities identified with photometric redshifts (Spitler et al. 2012, Capak et al. 2011, Trenti et al. 2012). Some of these high redshift clusters have been confirmed spectroscopically (Papovich et al. 2010, Steidel et al. 2005, Tanaka et al. 2010 and Gobat et al. 2011). 

However, so far there has been no systematic analysis of high redshift groups in spectroscopic redshift surveys. As described below, the zCOSMOS-deep survey provides a large sample of galaxies at $z > 1$ including 3502 galaxies with usable redshifts in the redshift interval $1.8 < z < 3$ in a single fairly densely sampled region of sky (Lilly et al. 2007, Lilly et al. 2012 in prep.), allowing the application of the same sort of algorithm as has been used to identify groups at $z < 1$.  
 
The aim of this paper is to identify possible groups at $1.8<z<3$, based on a simple linking length algorithm.   We provide a catalogue of 42 such associations. In order to understand the physical nature of these detected structures, we have carried out extensive comparisons with mock catalogues that have been generated by Kitzbichler \& White (2007) and then passed through the same ``group-finding'' algorithms.  The primary aim is to assess whether the galaxies in these structures are indeed already occupying the same dark matter halo. We can however also use the mocks to follow the future fate of each galaxy and thus to see when, if ever, the candidate member galaxies will be in the same halo, whether they will merge with other galaxies and so on, and what the structures identified at high redshift are likely to become by the present epoch. 
 
This paper is organized as follows: We first describe the zCOSMOS-deep sample and the mock catalogues used to calibrate and analyze our group catalogue. In section 3 we develop our group-finder algorithm on the basis of comparisons with the mocks, and produce the catalogue of 42 associations. In Section 4 we carry out an extensive analysis of the mocks to see what they indicate for (a) the nature of the structures that we detect at $z \gtrsim 2$, (b) how they develop over time, down to $z \sim 0$, and © how representative they are of the population of progenitors of massive haloes today. In Section 5 we examine a complementary photo-$z$ sample and identify a significant excess of massive galaxies in the regions of the groups, but do not find evidence for any color differentiation of the population relative to the field, although we argue we should probably not have expected to see such differentiation.  We then conclude the paper and summarize our findings. 

Where needed we adopt the following cosmological parameters (consistent with the Millennium simulation): $\Omega_{m}=0.25, \Omega_{\Lambda}=0.75$ and H$_{0}=73$\,km\,s$^{-1}$\,Mpc$^{-1}$. All magnitudes are quoted in the AB system.

\section{Data}

\subsection{The zCOSMOS-deep sample}

The zCOSMOS-deep redshift survey (Lilly et al. 2007, Lilly et al. 2012 in prep.) has observed around 10'000 galaxies in the central $\sim$1\,deg$^{2}$ of  the COSMOS field. The selection of the targets for zCOSMOS-deep was quite complicated.   All objects were color-selected to preferentially lie at high redshifts, through (mostly) a $BzK$ color selection (c.f. Daddi et al. 2004) with a nominal $K_{\mathrm{AB}}$ cut at 23.5, supplemented by the purely ultraviolet $ugr$ selection (c.f. Steidel et al. 2004).  An additional blue magnitude selection was adopted that for most objects was $B_{\mathrm{AB}} < 25.25$.   These selection criteria yield a set of star-forming galaxies which lie mostly in the redshift range $1.3 < z < 3$ (Lilly et al. 2007).
The targeted sources were then observed with the VIMOS spectrograph at the VLT using the low resolution LR-Blue grism giving a spectral resolution of $R = 180$ over a spectral range of $3700 - 6700$ \AA.   The spatial sampling of zCOSMOS-deep is such that a central region of $0.6^{\circ}\times0.62^{\circ}$ was covered at approximately 67\% sampling, with a lower sampled outer region extending out to $0.92^{\circ}\times0.91^{\circ}$. Both regions are centered on 10 00 43 (RA) , 02 10 23 (DEC) .\\
In total 9523 galaxies have been observed. It was possible to assign a spectroscopic redshift to 7773 of them.  Repeat observations, including some with the higher resolution FORS-2 spectrograph indicate a typical velocity error of around 300\,km/s in the redshifts. 

To account for the varying reliability of the assigned spectroscopic redshifts, confidence classes have been introduced as described in detail in Lilly et al. (2009, 2012 in prep.). Objects with flags 3 and 4 have very secure redshifts, whereas objects with flags 1 and 2 have less secure redshifts. Flag 9 indicates a single narrow emission line.  An additional decimal place is used to  indicate the agreement with the photometric redshift, putting 0.5 if $|z_{\mathrm{phot}}-z_{\mathrm{spec}}|< 0.1(1+z)$, which is approximately three standard deviations of the scatter between photometric and spectroscopic redshifts.  

In this paper we only use galaxies with flags 3, 4, 1.5, 2.5 and 9.5 meaning that the corresponding redshifts are either secure on their own or confirmed by the respective photometric redshifts. Furthermore, we restrict our analysis to the redshift range $1.8 < z < 3$ where the success rate in measuring secure redshifts is highest because of the entrance of strong ultraviolet absorption features into the spectral range. The final sample used in this paper consists of 3502 objects from the catalogue in Lilly et al. (2012, in prep.).  In the central $0.36$\,deg$^2$ region the overall sampling rate of this sample relative to the target catalogue is about 55\%. We have a comoving number density of $6.1 \times 10^{-4}$\,Mpc$^{-3}$.

\subsection{Mock catalogues}

\subsubsection{The Millennium Simulation}
The Millennium Simulation is a large dark matter $N$-body simulation carried out in a cubic box of 500\,h$^{-1}$\,Mpc sidelength. It starts from a glass-like distribution of particles that is perturbed by a gaussian random field and it follows the evolution of dark matter particles from $z=127$ to $z=0$. The results are stored in 64 snapshots, placed logarithmically in redshift space and starting from $z=20$. From these dark matter particles merger-trees are built up through the identification of gravitationally bound haloes which in post-processing are populated with galaxies  (Springel et al. 2005, Lemson \& Springel 2006). Several semi-analytic models for the galaxy formation process have been implemented on top of the dark matter structure of the Millennium simulation. The Kitzbichler \& White (2007) mocks used in this work are based on a galaxy formation semi-analytic model (SAM) as described in deLucia \& Blaizot (2007). 

The structure and presentation of the Millennium simulation allows us to follow both haloes and individual galaxies through time and therefore to determine the subsequent evolution of group-like structures that are identified at a particular redshift (Lemson et al. 2006). It is therefore ideal for the present purposes of trying to understand the physical nature of corresponding objects in the sky, provided of course that the simulation, and the associated galaxy formation model, are not grossly inconsistent with the real Universe.  

In this work we make extensive use of the six independent \citet{kitzbichler07} mock lightcones which provide ``observations'' of a $1.4^{\circ} \times 1.4^{\circ}$ field and in which the identities of the galaxies are linked to the Millennium Simulation.  These light cones are constructed with an observer at redshift $z = 0$ using a periodic extension of the simulation box to cover high redshifts (Blaizot et al. 2005).  This will inevitably lead to the eventual double appearance of objects. However, for the field size of $1.4^{\circ} \times 1.4^{\circ}$ the first duplicate will appear around $z\sim5$, which is beyond the redshift range we are interested in. Each light cone is based on a different observer and a different direction and therefore can be regarded as independent in terms of large scale structure at high redshifts. 
The mocks give the positions of galaxies in RA and DEC, as well as the observed redshifts, including the effects of peculiar velocities.

\subsubsection{Sample selection}

For the mock catalogues to resemble the zCOSMOS-deep sample we first add a straightforward observational velocity error to each galaxy by adding a velocity selected randomly from a gaussian distribution with $\sigma_{v}=300\,$km/s.  The main concern is to match the number densities of galaxies in the actual zCOSMOS sample and in the mocks.  Starting with the set of all galaxies in the mocks, we applied limiting magnitudes in $B$ and $K$.  Small adjustments to the nominal $B_{\mathrm{AB}} < 25.25$ and $K_{\mathrm{AB}} < 23.5$ limits were then made above and below $z \sim 2$ so as to match as well as possible the shape of the $N(z)$ number counts of objects in the actual data, i.e., so that $s=\Sigma_{\mathrm{mocks}} \Sigma_{z} (N_{\mathrm{data}}(z)-N_{\mathrm{mocks}}(z))^2$ was minimized.  Given the overall sampling (spatial sampling times spectroscopic success rate) of zCOSMOS-deep in this redshift range, we constructed, through these small magnitude adjustments, a mock sample that had exactly twice the surface number density as the final spectroscopic sample in the highly sampled central region. This meant a final division of the mock sample into two via random sampling could be used to simulate the $\sim$50\% sampling of the spectroscopic data and yield a second, complementary, mock sample from the same light cone. This is useful to see the effects of the sampling as well as doubling the number of mock samples. 

\begin{figure}[t!]
\includegraphics[scale=0.41]{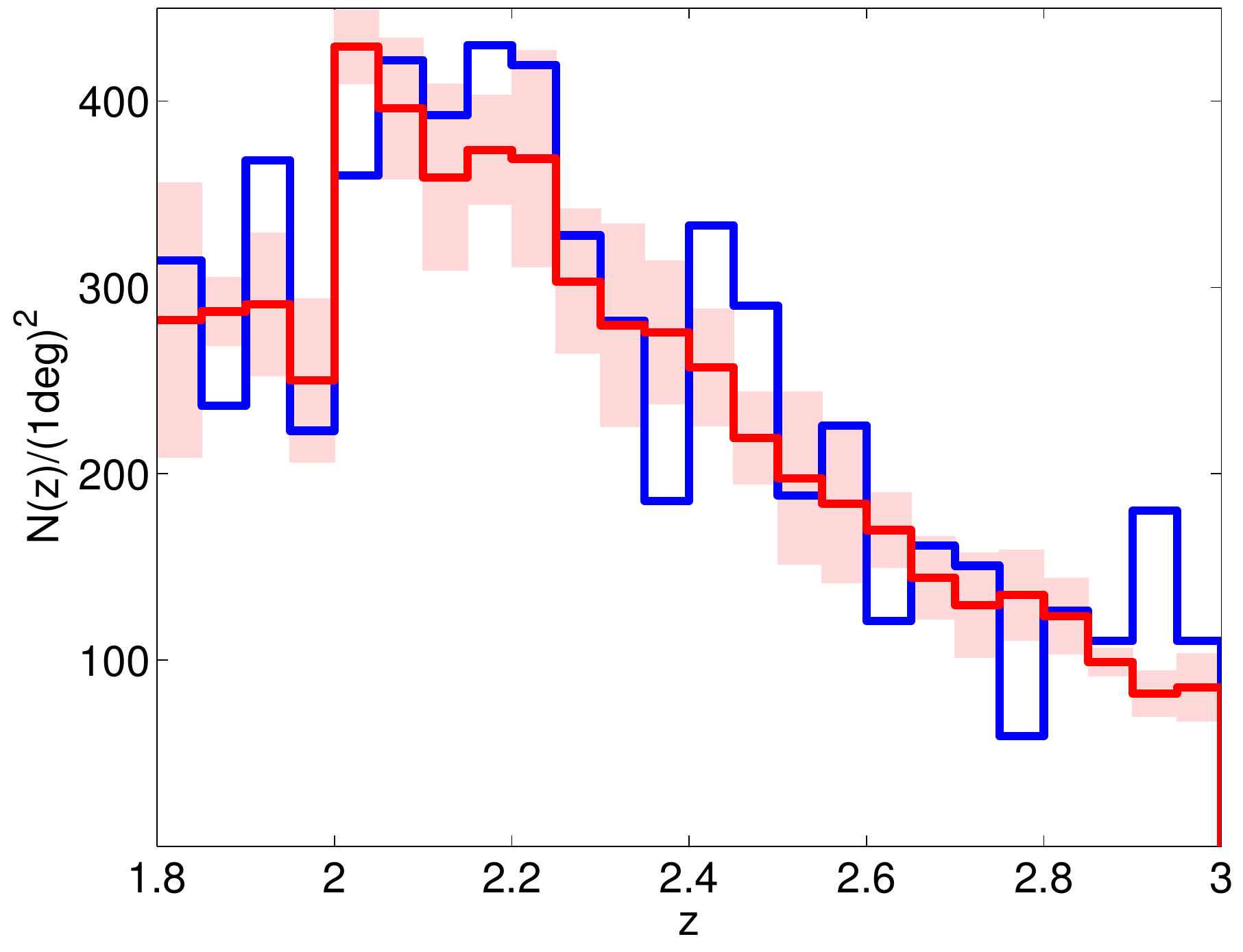}
\caption{\small The average $N(z)$-distribution of the objects in the final mock catalogues (red) after adjustment, as compared to the $N(z)$-distribution of the actual zCOSMOS-deep sample (blue). The shaded area shows the spread of the mocks (in terms of their standard deviation). An adjustable magnitude cut in $B$ and $K$ was applied to the mocks in order to match the number density of galaxies to the data (see text). \normalsize}
\label{fig_sfrAndNz}
\end{figure}

It should be emphasized that the goal of this exercise was to produce a mock sample that had the correct $N(z)$ and was similarly dominated by star-forming galaxies (by making similar nominal cuts in B and K as in the zCOSMOS selection), rather than to simulate exactly the selection of the objects.  Such an exact simulation would have depended on the details of the galaxy formation prescription used in the SAM prescription, and on the uncertain vagaries of the zCOSMOS-deep spectroscopic success rate etc.   Figure \ref{fig_sfrAndNz} shows the resulting $N(z)$ averaged over all twelve mock samples, compared with that of the zCOSMOS sample.

\section{Methods}

\subsection{Group definition}

Throughout this work we will use the following terminology:
\begin{enumerate}
\item ``(real) group'': a set of three or more galaxies which are all in the same dark matter halo at the epoch in question;
\item ``partial group'': a set of three or more galaxies at least two of which are in the same dark 

\begin{figure}[h!]
\includegraphics[scale=0.5]{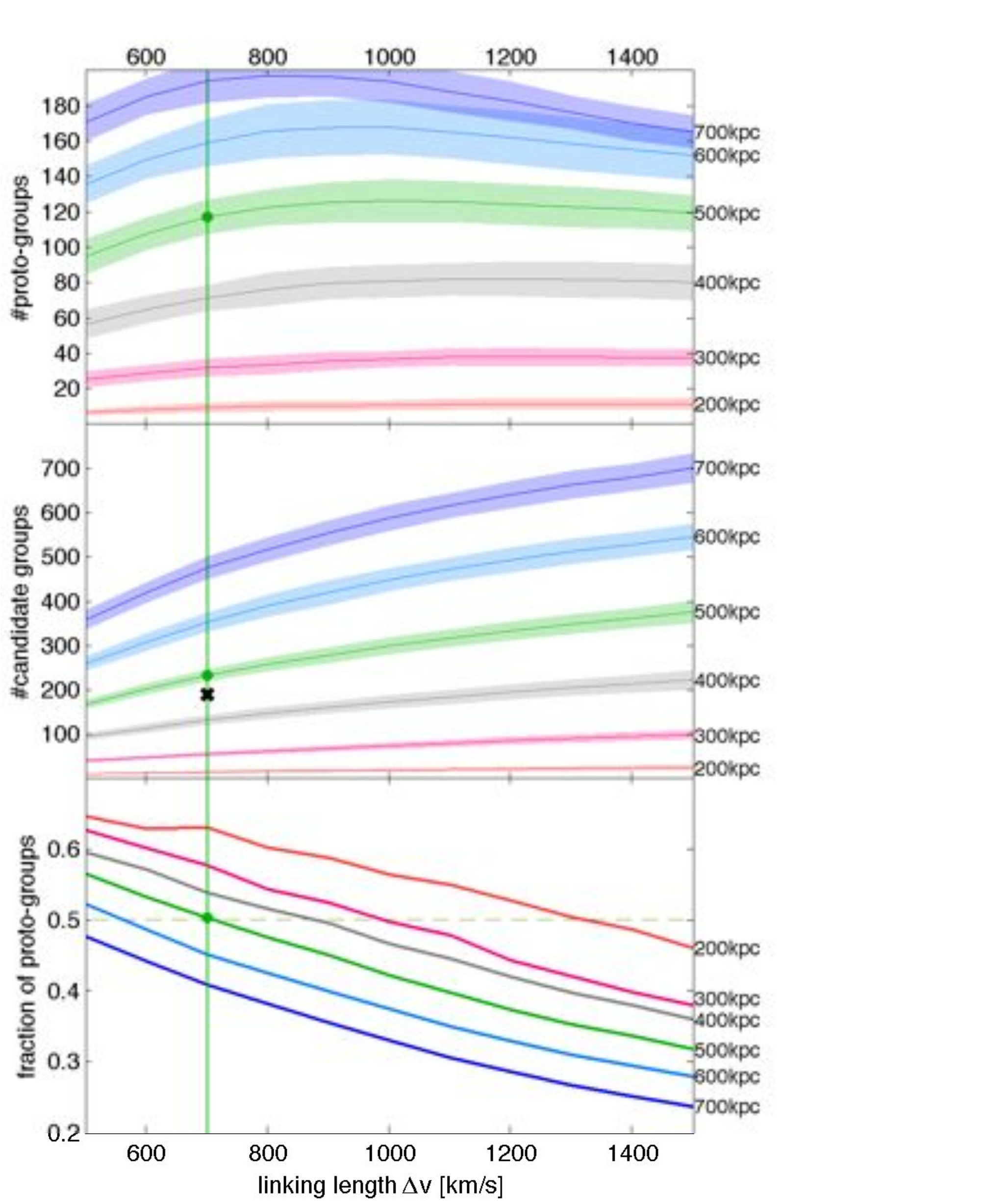}
\caption{\small Number of proto-groups at $1.8<z<3$, which includes any real at this redshift, in the mock catalogues (upper panel), the total number of candidate groups (middle) and the fraction of (proto-) groups (the fraction of the detected structures which either already constitute a group or will do so by $z=0$, lower panel) as a function of the velocity linking length $\Delta v$ for various projected linking lengths $\Delta r$. The numbers show the average number per mock catalogue,   and the shaded areas the spread in the mocks in terms of their standard deviation. The number of (proto-)groups stays largely constant after the first rise up to $\Delta v\sim700$\,km/s, whereas the total number of candidate groups keeps rising with increasing $\Delta r$ and $\Delta v$, producing a declining fraction of (proto-)groups.  Requiring the velocity linking length to fulfill $\Delta v\gtrsim 700$\,km/s, and the choice of 500\,kpc for the projected linking length (shown in green) keeps the fraction of proto-groups above 50\% (see text for details). The middle panel also shows the actual number of candidate groups found in zCOSMOS-deep with this parameter choice (black cross).  This is in good agreement with the number of candidate groups defined in the same way in the mock catalogues. \normalsize}
\label{fig_varLL}
\end{figure}
\clearpage

matter halo at the epoch in question;
\item ``candidate group'': a set of three or more galaxies that are identified by the group-finder as defined in the next section;
\item ``proto-group'': a candidate group in which all the members will be found in a real group at some later epoch;
\item ``partial proto-group'': a candidate group which will become a partial group at a later epoch, i.e., in which some apparent members at the epoch in question will never appear in the same halo down to $z=0$;
\item ``spurious group'':  a candidate group in which none of the apparent members will ever belong to the same halo down to $z = 0$, i.e., the galaxies are simply projected on the sky.
\end{enumerate}

\subsection{The nature of groups in the mocks}

The Kitzbichler light cones provide the galaxies together with a link to the actual object within the Millennium simulation. Dark matter haloes are identified within the Millennium simulation using a friends-of-friends (FOF) algorithm applied to the dark matter particles. Galaxies belonging in the same halo therefore have the same halo identification number (FOF-ID) at the epoch in question (Lemson et al. 2006).  By examining, at all later times, the halo FOF-IDs of the galaxies which we have placed in candidate groups at $z \sim 2$, we can see when, if ever, these galaxies belong to the same halo. This makes it straightforward to determine the group nature (as defined above) of a particular set of galaxies that has been detected by application of the group-finder algorithm to a mock catalogue simulating an observational light cone. The galaxies in a proto-group will not share the same FOF-ID until the galaxies have entered the common halo. 

In our analysis, we have not considered the effect of changing the dark matter linking length in the Millennium simulation. For a discussion see Jenkins et al. (2001).

Likewise, the descendant tree of galaxies that is provided by the Millennium simulation can be used to follow the evolution of single galaxies from $z\sim2$ to $z=0$ and thereby to identify mergers between galaxies. When two galaxies have the same descendant at the next snapshot, they must have merged in the intervening time.

Using the mocks and the descendant trees of galaxies we were therefore able to identify, in the mocks, which candidate groups are already real or partial groups, which are not yet real/partial but will become so at some point in the future, and which are totally spurious in that the galaxies will never reside in the same halo.  We can also see which galaxies merge together, which by definition requires them to be in the same halo.

\subsection{Group finder algorithm}

There is an extensive literature on finding groups in spectroscopic redshift surveys, based on a friends-of-friends approach (Huchra \& Geller 1982, Eke et al. 2004, Berlind et al. 2006), the Voronoi Delaunay method (Marinoni et al. 2002, Gerke et al. 2005, Cucciati et al. 2010), or a combination of both (Knobel et al. 2009 and 2012). At lower redshifts, where the emphasis is on real groups in the same halo, the group finder should ideally only pick out real groups, minimizing the number of interlopers. A major concern is the over-merging or fragmentation of groups and a great deal of effort goes into controlling these issues (see \citet{knobel12} for an extensive discussion).   Many group-finders use a friends-of-friends method to link galaxies into structures. In choosing the linking lengths $\Delta r$ (in physical space) and $\Delta v$ one has to take into consideration the following, sometimes contradicting, requirements:
\begin{itemize}
\item The linking length has to be large enough to ideally encompass all groups that are present, but small enough for not to overmerge groups, i.e., miss-detect two distinct groups as one.

\begin{figure}[t!]
\includegraphics[scale=0.42]{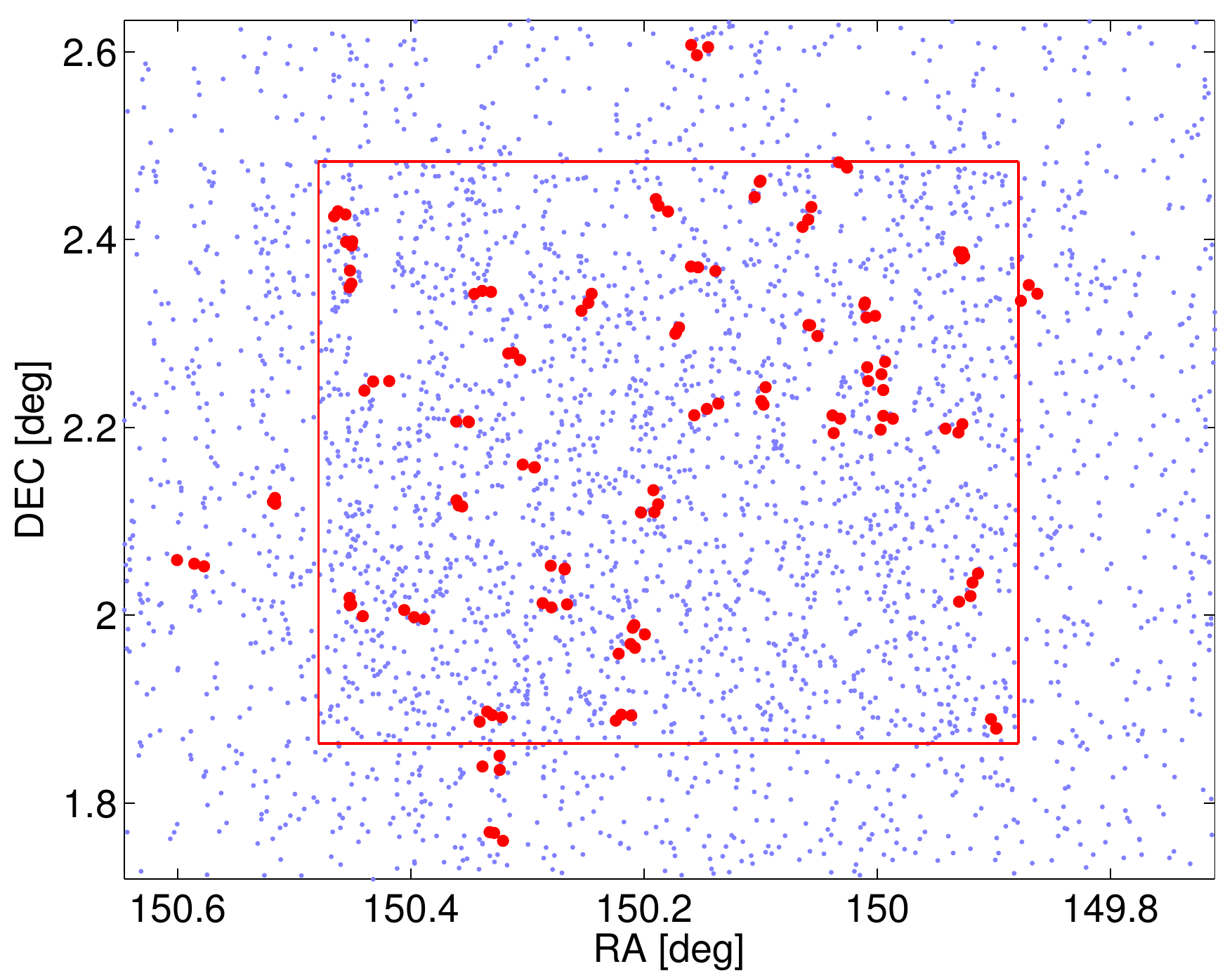}
\caption{\small The location of candidate groups in the COSMOS field.  The candidate members are shown in red.  The underlying zCOSMOS-deep sample in the same redshift range is shown in blue.  The red square shows the extent of the central, highly sampled, area.  Not surprisingly, the detection of structure is sensitive to the projected density of the available tracers.\normalsize}
\label{fig_locationGroupsDeep}
\end{figure}

\item Interlopers (i.e., miss-identified group galaxies) should be avoided.
\item The linking lengths must take into account the measurement errors as well as peculiar velocities.
\end{itemize}

The choice of values for the linking lengths is therefore a compromise.  We explored the performance of the group-finder with varying linking lengths with the mock catalogues, determining for each resulting group catalogue the total number of candidate groups, the total number of real groups plus proto-groups, and the fraction of real (proto-)groups, i.e., the fraction of the detected structures at $z\sim2$ which either constitute a group already then, or will do so by $z=0$. This is shown as a function of linking length in Figure \ref{fig_varLL}.
It turns out that the number of real (proto-)groups stays largely constant with increasing velocity linking length beyond $\sim700$\,km/s, but increases with linking length $\Delta r$. 
The total number of candidate groups however increases steadily with both $\Delta v$ and $\Delta r$, meaning that the fraction of real (proto-)groups decreases with $\Delta v$ and with $\Delta r$.  We set a fraction of real (proto-)groups of 50\% as a minimum requirement. The remaining 50\% of the sample will contain a significant number of partial (proto-)groups, which will increase the success rate  (see section 4.1). Because of the initial upturn in the number of real (proto-)groups we also want to have $\Delta v\gtrsim 700$\,km/s. It then turns out that the maximal linking length $\Delta r$ (physical space) that fulfills these two requirements is 500 kpc. The $\Delta r=500$\,kpc and $\Delta v=700$\,km/s are slightly higher values than for instance in \citet{knobel09}, who uses 300-400\,kpc and $\sim$ 400\,km/s. This is, however, justified by the larger measurement errors at our higher redshifts and the lower density of our tracer galaxies.

The width of the shaded area in the two upper panels of Figure \ref{fig_varLL} indicates the standard deviation in the number of proto-groups and candidate groups in the 12 mocks. This shows that cosmic variance is small compared to Poisson noise.

\begin{figure}[t!]
\includegraphics[scale=0.42]{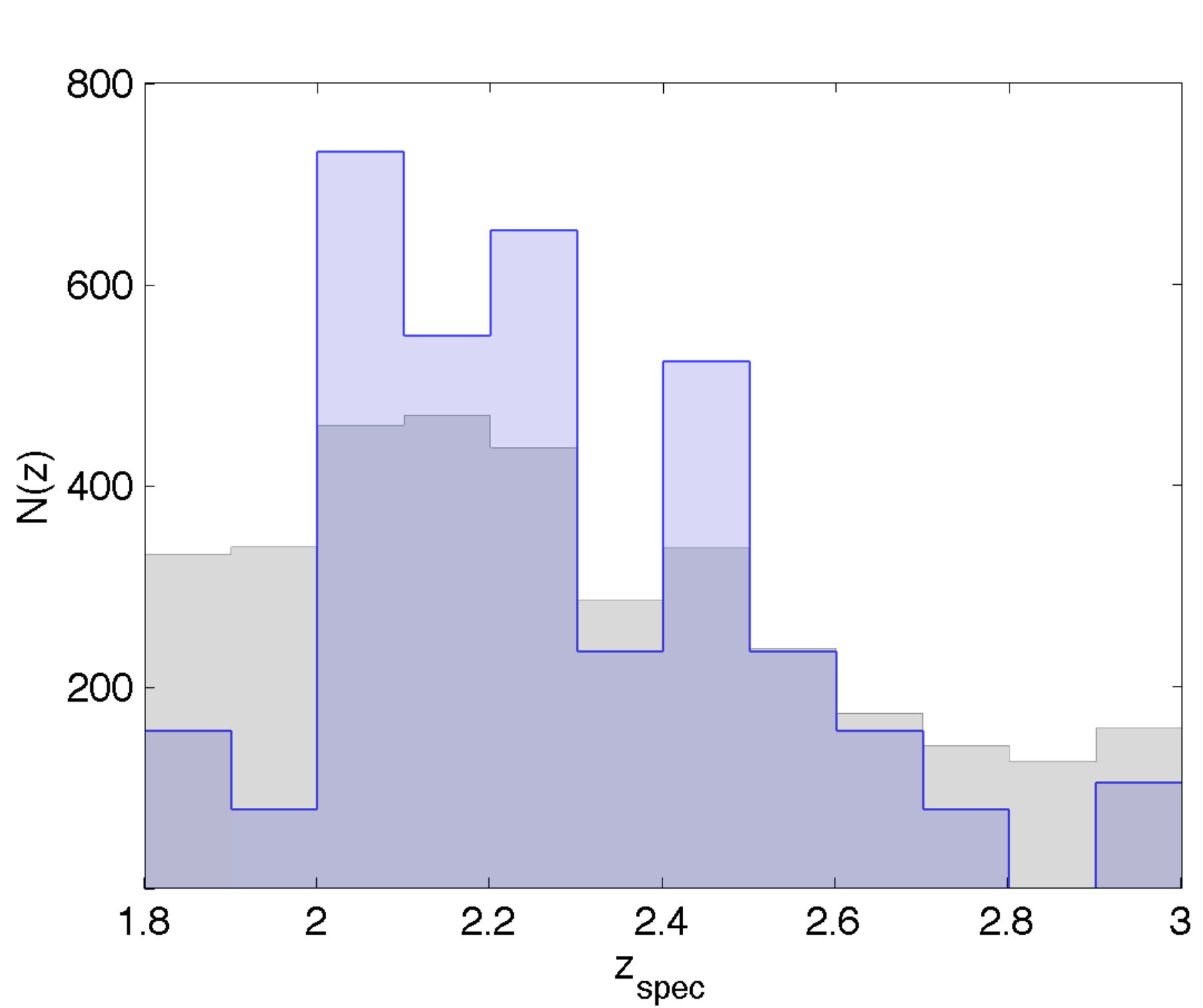}
\caption{\small The $N(z)$-distribution of the galaxies in the actual zCOSMOS-deep candidate groups (blue) compared to the distribution of the whole sample (grey), normalized to the 
same number of galaxies.\normalsize}
\label{fig_lNzGroupsDeep}
\end{figure}

\subsection{Application to zCOSMOS sample and comparison with mocks}

 Having determined the parameters of the FOF algorithm in the previous section, we apply the group-finder to the actual zCOSMOS data and the 12 mock samples. In the data this results in 42 candidate groups with memberships of three or more, i.e., we do not consider ``pairs''. Of these 42, one has five members and six have four, so the vast majority are triplets. The 42 candidate groups are listed in Table 1, their redshift distribution as compared to the parent sample is shown in Figure \ref{fig_lNzGroupsDeep}. Almost all of the detected candidate groups are in the central more highly sampled region of the field, as shown in Figure \ref{fig_locationGroupsDeep}.

For each zCOSMOS candidate group, and for the corresponding candidate groups in the mocks,

\onecolumn
\begin{tabular}[b!]{|c|c|c|c|c|c|c|c|c|}
\hline
ID & ID of one member & $<RA>$ & $<DEC>$ & $<z>$ & $r_{\mathrm{rms}}$ [kpc] & $v_{\mathrm{rms}}$ [km/s]  & Richness\\
\hline
30 & 431260 & 150.151 & 2.369 & 2.463 & 325 & 30 & 3 \\
9 & 426916 & 150.278 & 2.011 & 2.308 & 322 & 87 & 3 \\
23 & 430182 & 150.312 & 2.277 & 2.578 & 193 & 94 & 3 \\
20 & 429414 & 150.147 & 2.219 & 2.090 & 362 & 101 & 3 \\
21 & 409768 & 150.43 & 2.246 & 2.157 & 366 & 104 & 3 \\
25 & 410733 & 150.172 & 2.302 & 2.099 & 117 & 112 & 3 \\
16 & 490781 & 150.297 & 2.158 & 2.099 & 185 & 130 & 3 \\
6 & 426643 & 150.206 & 1.985 & 2.232 & 229 & 140 & 3 \\
7 & 426726 & 150.397 & 2.000 & 2.707 & 287 & 143 & 3 \\
19 & 429340 & 149.993 & 2.206 & 2.554 & 279 & 147 & 3 \\
39 & 429401 & 150.036 & 2.205 & 2.096 & 324 & 206 & 3 \\
42 & 434564 & 149.870 & 2.343 & 2.678 & 319 & 222 & 3 \\
5 & 426418 & 150.214 & 1.964 & 2.117 & 269 & 227 & 3 \\
17 & 429152 & 149.933 & 2.199 & 2.279 & 261 & 239 & 3 \\
26 & 411468 & 150.249 & 2.333 & 2.469 & 297 & 239 & 3 \\
13 & 407675 & 150.194 & 2.118 & 2.178 & 385 & 251 & 4 \\
32 & 434605 & 150.452 & 2.396 & 2.286 & 110 & 254 & 3 \\
36 & 413529 & 150.102 & 2.456 & 2.476 & 294 & 264 & 3 \\
28 & 411517 & 150.338 & 2.344 & 1.805 & 224 & 281 & 3 \\
40 & 429794 & 150.098 & 2.232 & 2.099 & 302 & 284 & 3 \\
35 & 413241 & 150.186 & 2.436 & 2.051 & 260 & 296 & 3 \\
41 & 434071 & 150.332 & 1.892 & 2.957 & 257 & 304 & 4 \\
34 & 431678 & 150.461 & 2.427 & 2.322 & 169 & 316 & 3 \\
2 & 402591 & 150.329 & 1.841 & 2.096 & 351 & 322 & 3 \\
11 & 427339 & 150.272 & 2.050 & 2.306 & 214 & 328 & 3 \\
12 & 406198 & 150.588 & 2.055 & 2.029 & 369 & 340 & 3 \\
10 & 490746 & 149.921 & 2.028 & 2.050 & 459 & 365 & 4 \\
29 & 431233 & 150.452 & 2.356 & 2.278 & 282 & 381 & 3 \\
1 & 424327 & 150.327 & 1.766 & 2.538 & 229 & 386 & 3 \\
14 & 428112 & 150.359 & 2.118 & 2.232 & 126 & 405 & 3 \\
3 & 425554 & 149.900 & 1.883 & 2.215 & 190 & 415 & 3 \\
4 & 425598 & 150.218 & 1.892 & 2.684 & 217 & 435 & 3 \\
27 & 430794 & 150.008 & 2.325 & 2.258 & 275 & 474 & 4 \\
37 & 413838 & 150.028 & 2.479 & 2.452 & 146 & 476 & 3 \\
33 & 413105 & 150.060 & 2.423 & 2.469 & 335 & 488 & 3 \\
38 & 433521 & 150.153 & 2.603 & 2.282 & 281 & 496 & 3 \\
8 & 426762 & 150.449 & 2.010 & 2.013 & 293 & 505 & 4 \\
15 & 428229 & 150.517 & 2.121 & 2.153 & 102 & 507 & 3 \\
18 & 420527 & 150.354 & 2.206 & 1.808 & 188 & 513 & 3 \\
22 & 430097 & 150.000 & 2.256 & 2.440 & 412 & 526 & 5 \\
31 & 431338 & 149.928 & 2.384 & 2.143 & 113 & 534 & 4 \\
24 & 410797 & 150.056 & 2.305 & 1.974 & 237 & 545 & 3 \\
\hline 
\end{tabular}
\vspace{5pt}

\textit{Table 1: Candidate groups detected in zCOSMOS-deep, ordered by their velocity dispersion $v_{r\mathrm{rms}}$ }

\twocolumn

\begin{figure}[t!]
\includegraphics[scale=0.41]{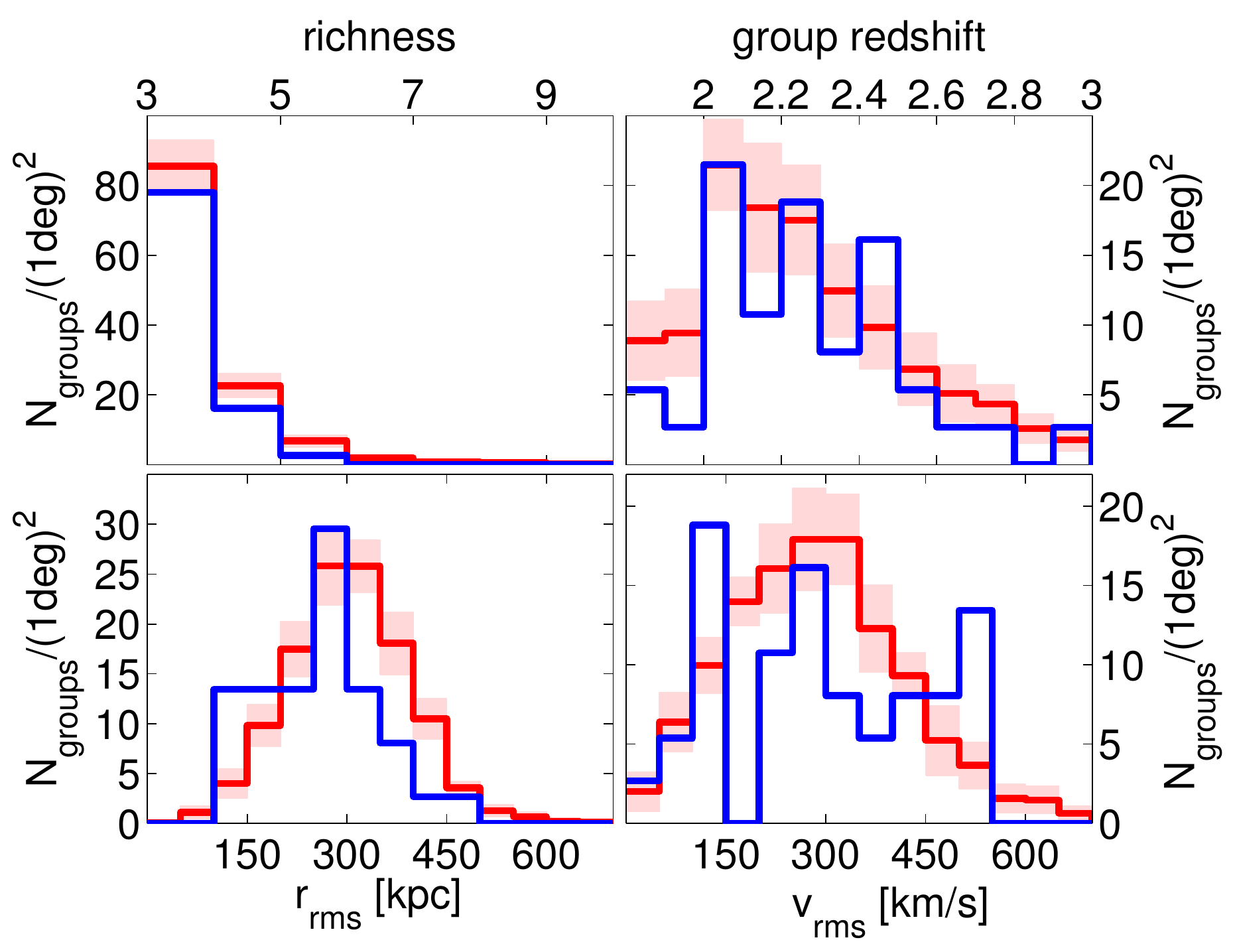}
\caption{\small Comparison of the basic properties of the candidate groups in the mock sample (red) with those in zCOSMOS-deep (blue). The shaded areas show the spread in the mock samples in terms of their standard deviation.
Top left: Richness (number of candidate member galaxies). Top right: Redshift of the candidate group. Bottom left: Root-mean-square radius of the candidate group, ($r_{\mathrm{rms}}$) defined as the r.m.s. distance of the members to their mean RA and DEC.  Bottom right: R.m.s. of the velocity  ($v_{\mathrm{rms}}$) relative to the center of the candidate group defined by the mean redshift of the members. 
In general there is a good agreement between mocks and data, in particular when taking into consideration the low number of candidate groups in the data.\normalsize}
\label{fig_deepVSmock}
\end{figure}
 
we also compute a nominal r.m.s. size and velocity dispersion by $r_{\mathrm{rms}} = \sqrt{\sum_{i} r_{i}^2/(N-1)}$ and
$v_{\mathrm{rms}} = \sqrt{\sum_{i} v_{i}^2/(N-1)}$,
where $r_{i}$ and $v_{i}$ denote the distance or the velocity of a galaxy to the center of the candidate group and $N$ is the number of members.

The center of the candidate group is defined by the average RA, DEC and $z$. The overall number of candidate groups found in the central area of zCOSMOS-deep (36 groups) agrees quite well with the average number found in the mocks, which is 44 per 0.36\,deg$^2$, i.e., the actual data has 18\% fewer candidate groups.   As shown in Figure \ref{fig_deepVSmock}, there is also broad agreement in the distributions in redshift, richness, and in the nominal size $r_{\mathrm{rms}}$ and velocity dispersion $v_{\mathrm{rms}}$ distributions.

\section{Results}

\begin{figure}[t!]
\includegraphics[scale=0.4]{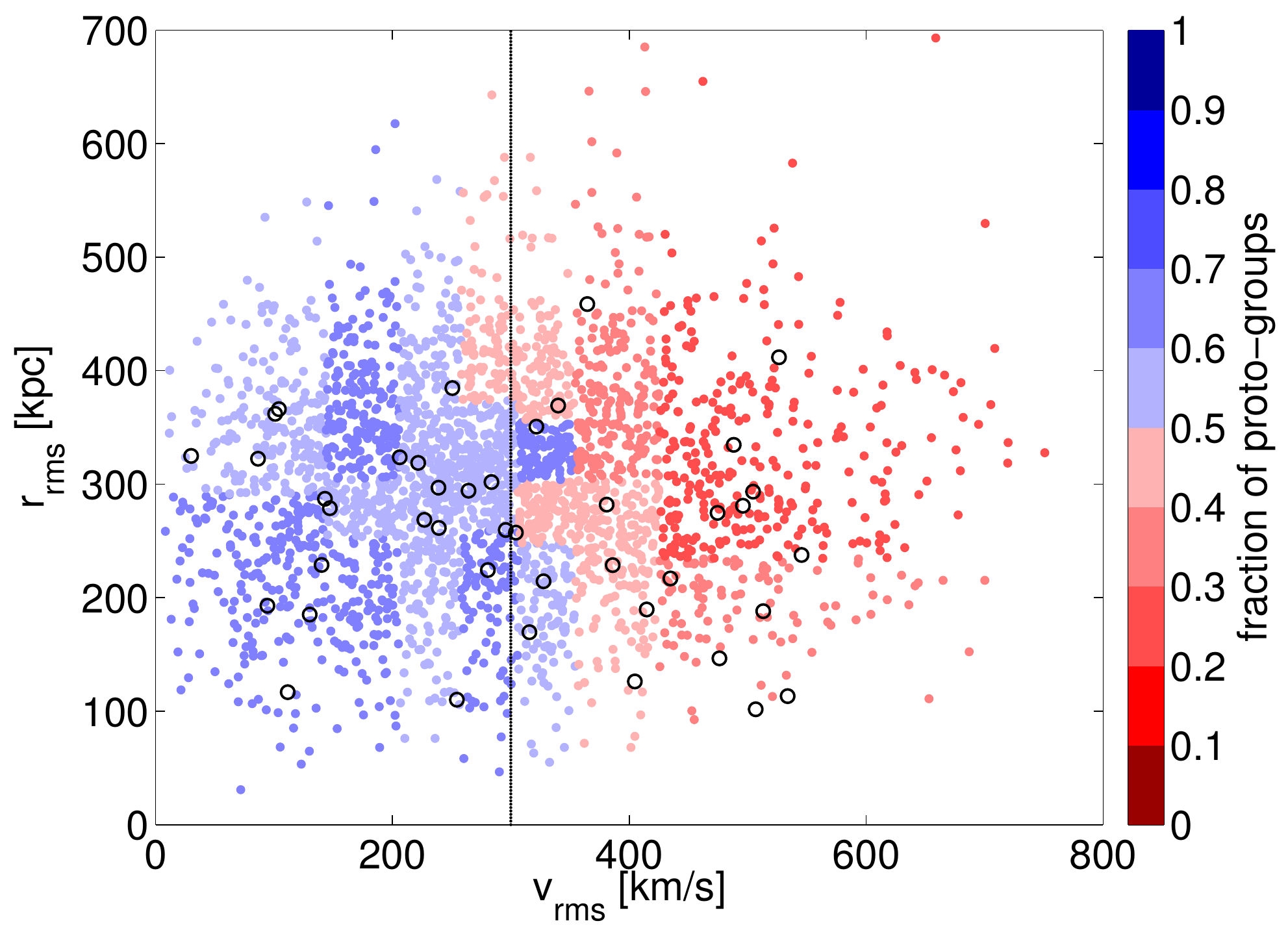}
\caption{\small The fraction of proto-groups with respect to all candidate groups in the mocks as a  function of their velocity dispersion $v_{\mathrm{rms}}$ and size $r_{\mathrm{rms}}$. This fraction strongly depends on $v_{\mathrm{rms}}$ whereas it is largely independent of $r_{\mathrm{rms}}$. For $v_{\mathrm{rms}}\lesssim 300$\,km/s the fraction of proto-groups is above 50\%. The observed $v_{\mathrm{rms}}$ is a crude indicator for the chance of a candidate group to become a real group in which all the galaxies share the same halo. The black circles show the location of the zCOSMOS-deep candidate groups. \normalsize}
\label{fig_2dvrmsrrms}
\end{figure}

\subsection{Are we detecting real groups at $z \gtrsim 2$?}

Whereas, as established in the following section, a significant number of the candidate groups will have assembled by $z=0$, we find that only 5 (out of in total 2791), i.e., less than 0.2\%, of the candidate groups in the mocks are real groups in the sense that all of the members are already in the same dark matter halo at the time of observation (i.e. at $z \sim 2$). However, $8\%$ of the observed structures are partially assembled with two galaxies in the same halo, meaning that we are observing groups with interlopers.  

The Millennium simulation used WMAP1 cosmological parameters (with a $\sigma_{8} = 0.9$), whereas the most recent cosmological data establish a lower value for $\sigma_{8}$, implying a lower build-up of structure at a given redshift.  As would be expected, the mock catalogues described in \citet{wang08}, where $\sigma_{8} = 0.81$ using the WMAP3 parameters (which are close to the most recent estimates) also yield essentially no real groups amongst the candidate groups at $1.8<z<3$.

\subsection{Assembly timescale}

\begin{figure}[t!]
\includegraphics[scale=0.41]{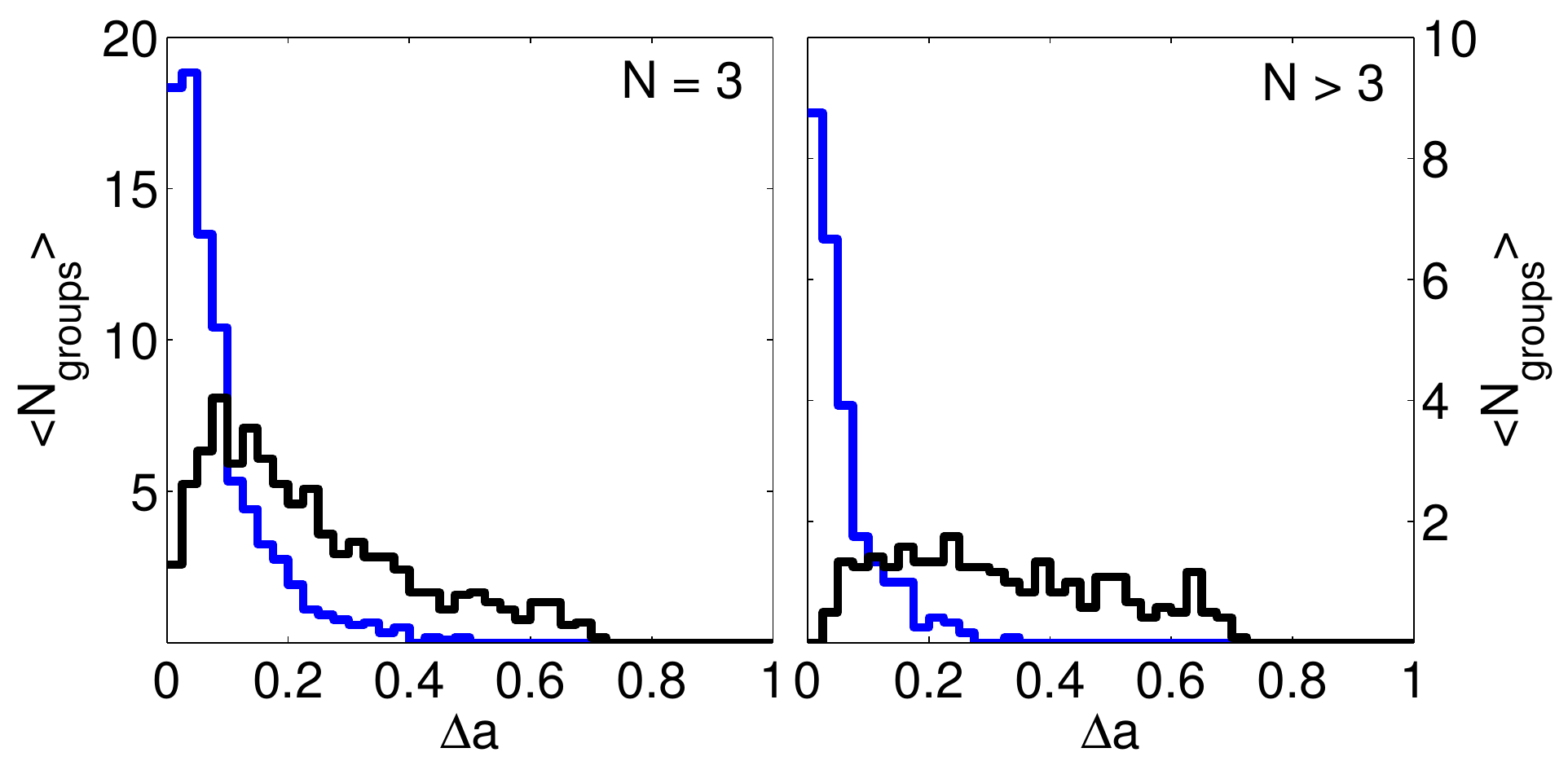}
\caption{\small The subsequent assembly of the proto-groups in the mocks.  The diagrams show the change in $a$, the cosmic scale factor, before the proto-groups have accreted two (blue), and then all (black), of their identified members into the same halo (left panel for richness 3, right panel for richness $\geq4$).  Most of the proto-groups observed at $1.8 < z < 3.0$ start to assemble within $\Delta a < 0.1$.\normalsize}
\label{fig_virtime}
\end{figure}

We established above, based on comparisons with the mocks, that most of the detected structures at $1.8<z<3$ have not yet assembled when we observe them. In 8\% of the mock candidate groups, two of the galaxies are already in the same halo, but essentially no candidate group has assembled all three members.  It is therefore an interesting question to see when and if these actually become groups, i.e., if they are what we call ``proto-groups'' at $z\sim2$. 

\begin{figure}[t!]
\includegraphics[scale=0.42]{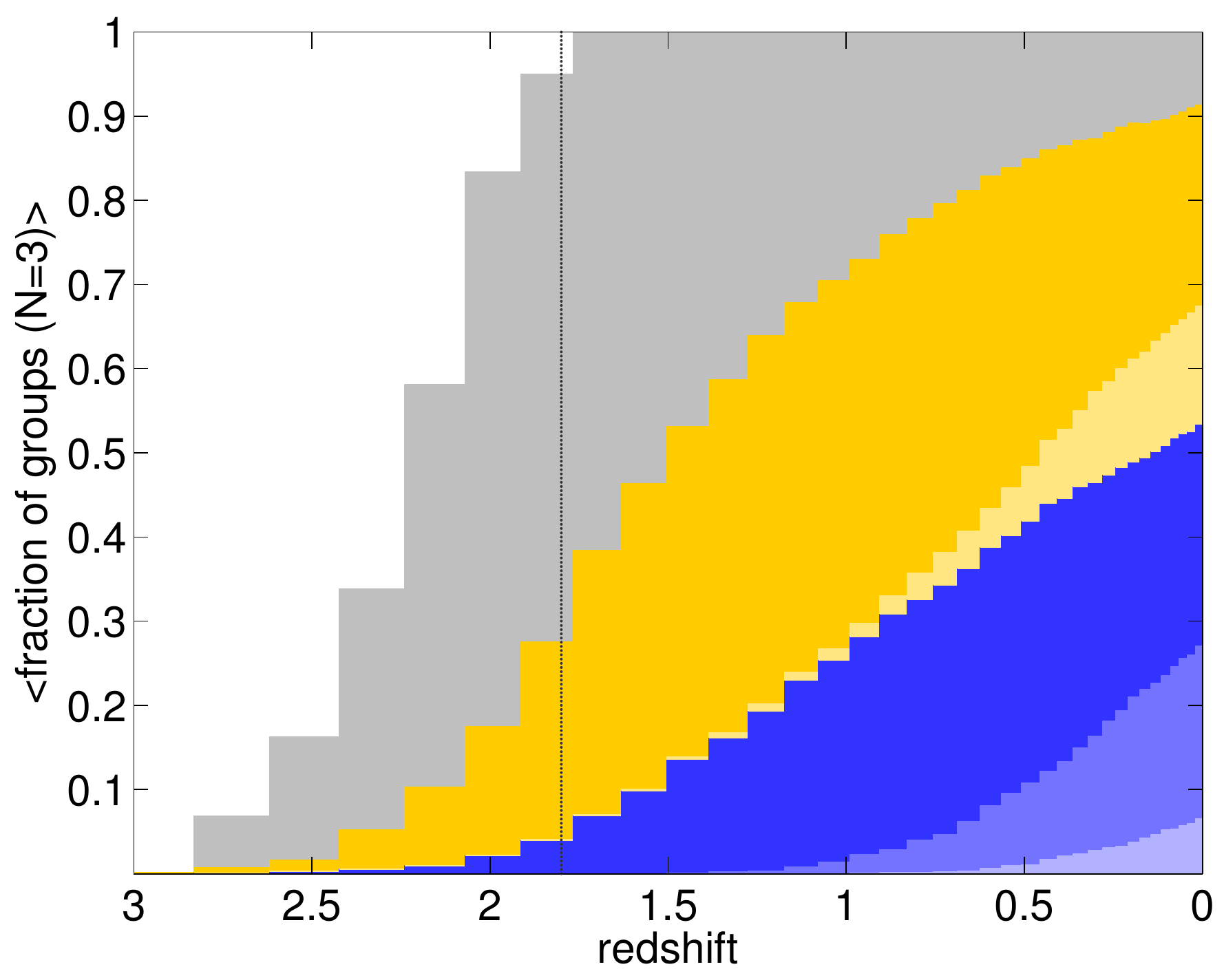}
\caption{\small The assembly history of all the candidate mock groups with richness 3 (which constitute over $\sim$85\% of the sample) over redshift. Partially assembled systems are shown in yellow (two members in the same dark matter halo) and fully assembled systems in blue (all members in the same halo). The light areas denote member galaxies that have subsequently merged (by definition within the same halo).  The grey zone represents candidate groups in which the members are not, at least yet, in the same halo.  The white zone is because we only follow the evolution of a candidate group after it has been detected in the light cone and the diagonal grey-white border therefore reflects the redshift distribution of the detected candidate groups. 
At $z=1.8$ already $\sim25\%$ of the candidate groups (detected at slightly higher redshifts) have assembled at least two of their members into the same DM halo, up from 8\% at the epoch of observation of the individual groups.\normalsize}
\label{fig_nsystems}
\end{figure}

The Millennium simulation allows us to follow the evolution of the structures we detect at $z\sim2$ down to $z = 0$, i.e., to see when, if ever, the structures detected in zCOSMOS will merge into a common halo.  It turns out that at the present time only 7\% of the detected mock candidate group galaxies are still completely outside of a common halo.  93\% of the mock candidate groups will either fully ($50\pm1$\%) or partially ($43\pm1$\%) assemble by the present epoch. The main criterion that distinguishes proto-groups from partial or spurious ones is the velocity dispersion $v_{\mathrm{rms}}$.  This is shown in Figure \ref{fig_2dvrmsrrms}.  In the regime $v_{\mathrm{rms}}\lesssim 300$\,km/s (which is comparable to the velocity error in the data) the fraction of mock proto-groups is above 50\%, whereas it drops below 50\% for velocity dispersions larger than 300\,km/s. The fraction of proto-groups does not depend on the projected radial size of the candidate group. The trend with velocity dispersion is, however, weak enough that it is not attractive to reject all candidate groups with $v_{\mathrm{rms}} \geq 300$\,km/s.

As stated above, 93\% of the mock candidate groups become real or partial groups by the present epoch. Already by $z \sim 1.5$, 50\% of the candidate groups are partial groups (up from 8\% at the epoch of observation, see \ref{fig_nsystems}) and by the current epoch, 50\% of the candidate groups at $z \sim 2$ are real groups with all detected members within the same halo. The majority of the proto-groups start to assemble within a $\Delta a < 0.1$ (see Figure \ref{fig_virtime}, ``a'' being the cosmic scale factor), which means that on a rather short timescale two or more members will share the same FOF-halo. The full assembly then requires a substantially larger timescale ($\Delta a\sim0.5$ or even more).

This continuous assembly process is further illustrated in Figure \ref{fig_nsystems} and emphasizes that assembly is taking place even within the observational ``window''.  Although only 8\% of the mock candidate groups are partially assembled by the time we observe them, by the end of the observing window at $z = 1.8$ around 25\% of the proto-groups have already members in the same dark matter halo.  These are therefore groups of richness 2 ``contaminated'' by an interloper (most of which obviously later on will accrete onto the group). According to the mocks, we are therefore able to actually observe the earliest phases of the assembly process of these groups.  

Figure \ref{fig_nsystems} also illustrates the likelihood that group members seen as distinct galaxies at $z \sim 2$ will have merged together by the current epoch.  In about 40\% of the proto-groups, two or more of the members that we identify at $z \sim 2$ will have merged together by the current epoch, and in about 10\% all three members will have merged into a single massive galaxy.

\subsection{Halo masses}

In the preceding discussion we followed the evolution of the structures that were detected by our group-finder at $z \sim 2$ down to the present epoch. In this section we look at haloes at the present epoch and ask which of their progenitors could have been detected at $z > 1.8$ in a zCOSMOS-like survey. To do this, we examine the set of all present-day haloes in the simulation whose progenitors lie within the $1.8 < z < 3$ volume of any of the six light cones. We first identify at the earlier epoch all of the haloes that will eventually assemble into a given present-day halo, and then identify all the ``progenitor galaxies'' within these progenitor haloes and ask if they satisfy the zCOSMOS brightness selection criteria, without the 50\% spatial sampling, referring to these as ``zCOSMOS-selected'' galaxies.
We then additionally ask whether this set of ``progenitor galaxies'' would have satisfied our group-funding requirements in terms of their spatial and velocity displacements, adding in also the incomplete spatial

\begin{figure}[htb!]
\includegraphics[scale=0.39]{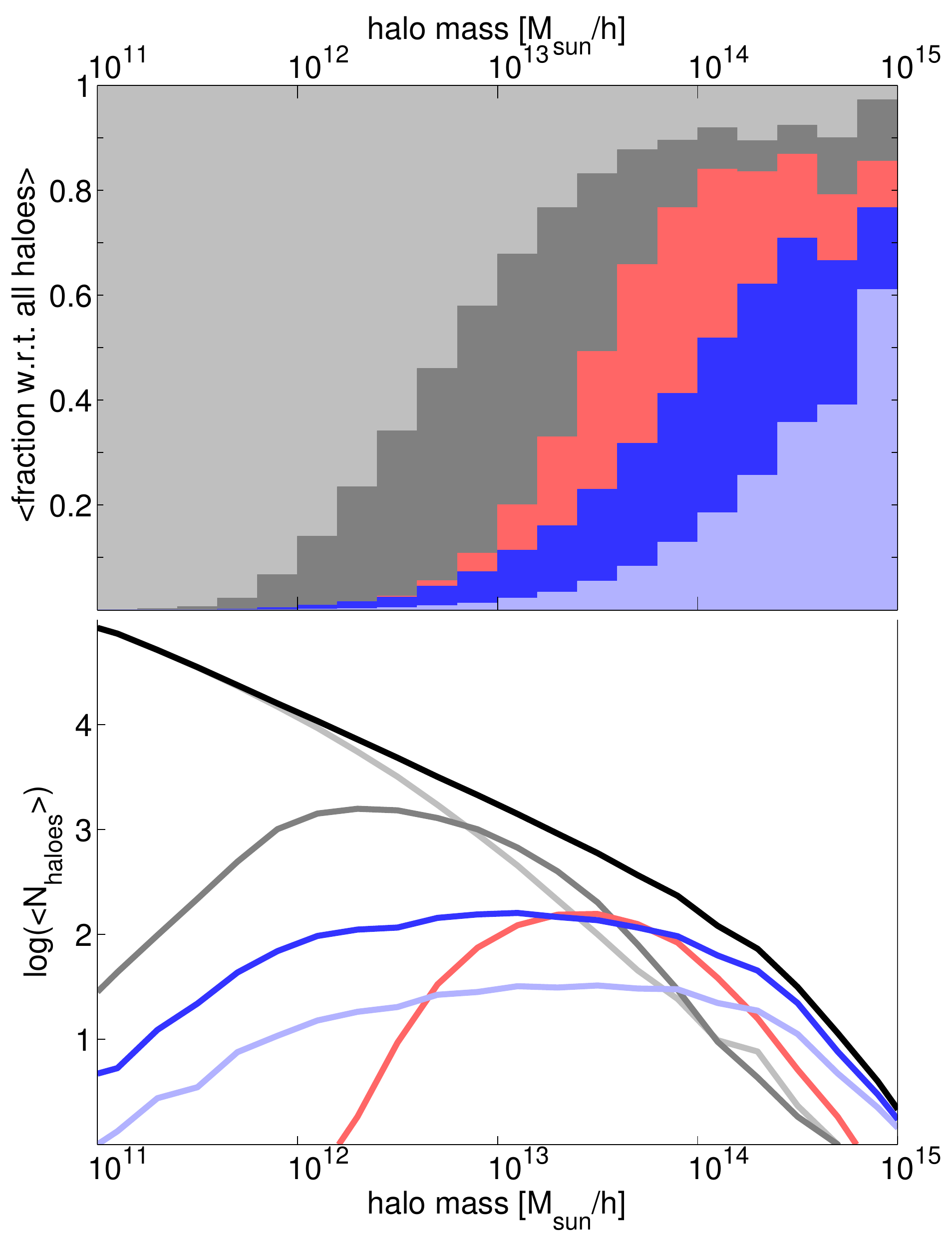}
\caption{\small Top panel: The average (over the 12 mock samples) fraction of present-day haloes that are detectable in a zCOSMOS-like survey at $1.8 < z < 3$, as a function of the present-day dark matter mass of the halo. 
The light blue region shows haloes which today contain three or more galaxies that, at high redshift, would satisfy the zCOSMOS-deep photometric selection criteria and would have been recognized as a candidate group with the zCOSMOS-deep overall sampling and success rates.  The dark blue region represents candidate groups that were not recognized simply because of the incomplete sampling/success rate - the lack of these in our candidate group catalogue was therefore simply a matter of chance. The pink region represents haloes in which the constituent galaxies would have been observed in zCOSMOS-deep, but which were too dispersed, in projected distance or velocity, to satisfy our group-finding algorithm.   The darker grey region represents present-day haloes which only had one or two progenitor galaxies satisfying the zCOSMOS-deep photometric criteria, while the light grey region represents haloes in which none of the progenitor galaxies could have been in zCOSMOS-deep.  Around 65\% of all present day $10^{14}-10^{15}$\,M$_{\odot}$/h groups would have a progenitor structure at $z\sim 2$ which we would in principle be able to identify in zCOSMOS-deep with full sampling.  Bottom Panel:  As in the upper panel, but now the the total number of haloes is plotted instead of the fraction.\normalsize
}
\label{fig_haloes}
\end{figure}
\clearpage

sampling of the zCOSMOS survey.

The result is shown in Figure \ref{fig_haloes}.   
Many haloes today, especially at $M < 10^{13}$\,M$_{\odot}$/h, do not have any zCOSMOS-selected progenitor galaxies at $1.8 < z < 3$. These are represented as the light grey region of the upper panel.   Some have only one or two zCOSMOS-selected progenitor galaxies and they are shown in dark grey, since they will by definition not be recognized as a ``proto-group''.  The pink region represents haloes today whose progenitor haloes did contain three or more zCOSMOS-selected galaxies but which were, at $1.8 < z < 3$, too dispersed to satisfy our group-finding linking lengths.  Finally, the blue region represents haloes with three or more progenitor galaxies that are close enough to be recognized as a candidate ``group''.  Applying the 50\% sampling of the zCOSMOS-survey, about a half of these are actually recognized (light blue), the remainder are missed simply because of the incomplete spatial sampling of the survey.

At high present-day halo masses (above $\sim 10^{14}$\,M$_{\odot}$/h) the majority of the haloes are represented in our candidate group catalogue in the sense of detecting three or more progenitor galaxies and recognizing them as members of a candidate group structure. In other words, around 65\% of todays $10^{14} - 10^{15}$\,M$_{\odot}$/h haloes should in principle have been recognized as a candidate group with the galaxy selection criteria of zCOSMOS, although a half of these will not have been detected in practice because of the random 50\% sampling of our survey.   Of the remaining 35\% of present-day haloes above $\sim 10^{14}$\,M$_{\odot}$/h that we would not have expected to be able to detect, more than a half have three or more detectable  
progenitor galaxies, but these are too dispersed in space or velocity to satisfy our criteria.   Increasing the linking lengths to catch these dispersed systems would, as shown above, however also severely increase the number of interlopers.

The lower panel of Figure \ref{fig_haloes} shows the distribtion of the present-day halo masses of the systems in our candidate group catalogue. While, as noted in the previous paragraphs, we are detecting a high fraction of the progenitors of the most massive haloes today, we are evidently  detecting a broad range of present-day halo masses with most systems in the $10^{13}-10^{14}$\,M$_{\odot}$/h range.

\subsection{Overdensities}

\subsubsection{Determination of the overdensity}

In order to give a rough estimate for the overdensities $\delta = \frac{\rho_{gr}-\bar{\rho}}{\bar{\rho}}$ associated with the candidate groups we calculated the mean 
(comoving) density  $\bar{\rho}$ of the overall sample in bins of $\Delta z = 0.2$ using the following equation: $$\bar{\rho} = \frac{N_{\Delta z}}{V}, V = \frac{1}{3}\cdot \mathrm{area} \cdot (l_{\mathrm{max}}^3-l_{\mathrm{min}}^3),$$ where $l$ denotes the comoving
distance along the line-of-sight and $area$ is the field of view of the mocks  (1.4$^{\circ}$x1.4$^{\circ}$).

The density of the groups $\rho_{\mathrm{gr}}$ was determined by assuming a cylinder with radius $r_{\mathrm{rms}}$ and a length of twice the $v_{\mathrm{rms}}$ (in comoving units):
$$\rho_{\mathrm{gr}} = 0.27 \cdot \frac{N}{\pi r_{\mathrm{rms}}^2 l} $$
where $N$ is the number of members, $l$ the length of the cylinder, and the factor 0.27 is included to account for the fact that in a 3D gaussian distribution only this fraction of the points would actually lie within the $1\sigma$ region (which we assumed here, by setting the size of the cylinder to the $r_{\mathrm{rms}}$ and the $v_{\mathrm{rms}}$).  

The overdensity computed here is at best a rough order of magnitude estimate. First, it refers to the density within the r.m.s. radius containing only a fraction of the observed galaxies, leading to an over-estimate of the mean overdensities of all of the galaxies in the structure.  An additional effect comes from the 50\% sampling rate. Adding in the missing galaxies does not add significant numbers of new members to the detected associations (since they were the lucky ones with above average sampling), whereas the mean density of the field increases by a factor of two, leading to a factor of up to two over-estimate in the overdensity. On the other hand, due the effect of  measurement errors in redshift (of order 300\,km/s) as well as peculiar velocities in that, the ``size'' along the line of sight may have been substantially over-estimated leading to an underestimate of the actual overdensity, e.g., by almost an order of magnitude since the observed $v_{\mathrm{rms}}$ corresponds to about 8\,Mpc (comoving) against the typical $r_{\mathrm{rms}}$ of $\sim1$\,Mpc (comoving). The estimated over densities should therefore be treated with considerable caution.

\subsubsection{Results} 

\begin{figure}[t!]
\includegraphics[scale=0.425]{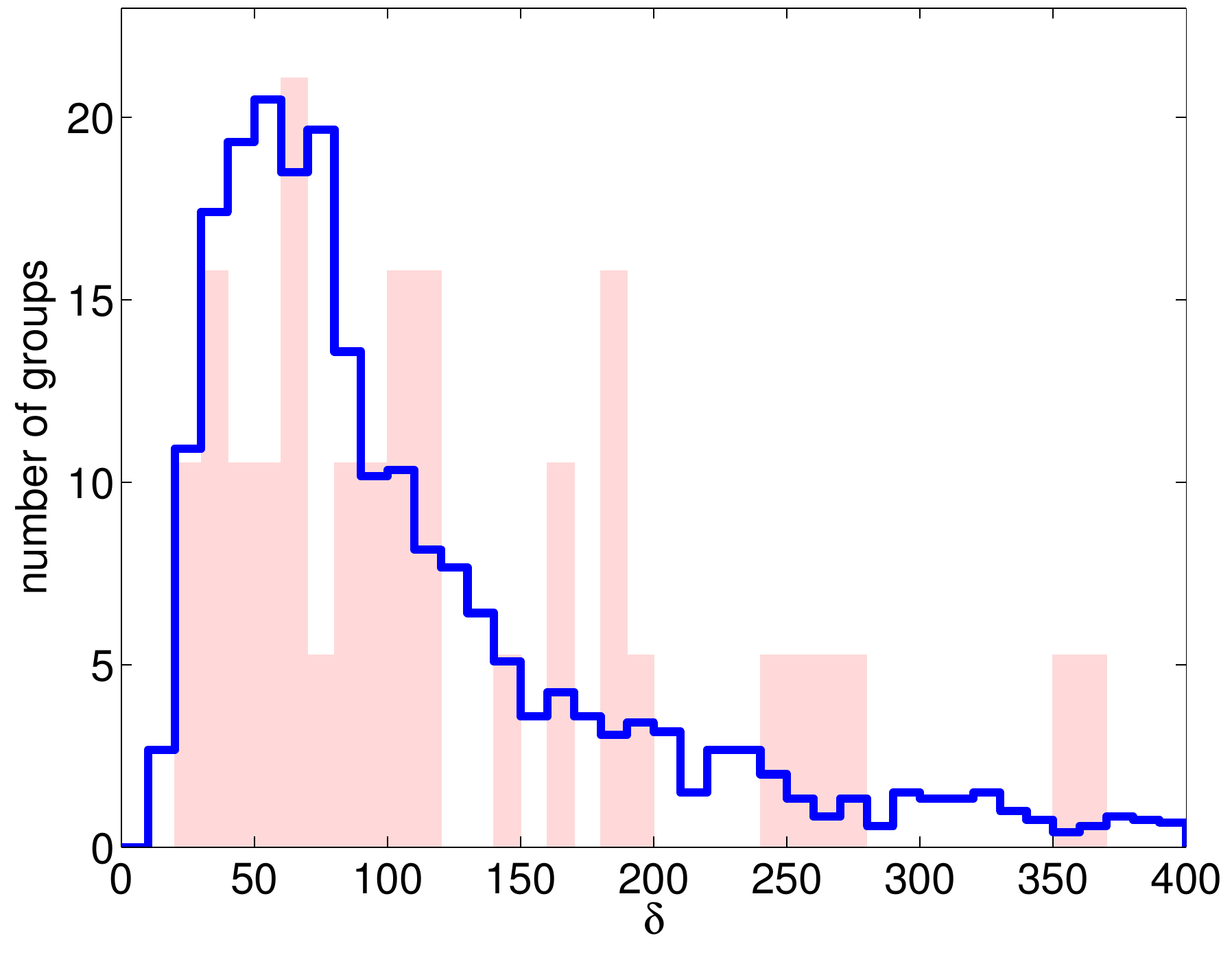}
\caption{\small The distribution of the group ``overdensities'' in zCOSMOS-deep (red) and in the mocks (blue).  These overdensities are quite large and indicate that the structures are in an advanced stage of collapse, consistent with the idea that the galaxies will assemble into the same haloes in the future.  However, readers should see the text for discussion and important caveats in the interpretation of this quantity.\normalsize}
\label{fig_deltaDistr}
\end{figure}

With these caveats in mind, the distribution of $\delta$ for the 42 candidate groups and for the corresponding mock samples is shown in Figure \ref{fig_deltaDistr}.  Even with the uncertainties outlined above, it is evident that that the candidate groups represent highly overdense regions and that most of them have probably already turned around (i.e., decoupled from the background).  This would be expected if they are to merge into a single halo within an interval of expansion factor of $\Delta a \sim a$ as discussed above. 

\subsection{Excess of high mass objects and red fractions}
So far we have established that the associations that we have found are in the main not yet fully formed groups, but are quite likely to become so by $z=0$. Furthermore, the candidate groups are already quite overdense.  For this reason it is of interest to look for surrounding overdensities and to look for any colour-differentiation of the galaxy population in and around the candidate groups relative to the field population.   Unfortunately, zCOSMOS-deep itself is limited to star-forming galaxies by the colour selection, and so it is necessary to use photo-$z$ objects from the larger and deeper COSMOS photometric sample (Capak et al. 2007).   Typical photo-$z$ errors are of order of $\Delta z \sim 0.03(1+z)$ or 10'000\,km/s.

We focus on relatively massive galaxies, above a stellar mass of $>10^{10}$\,M$_{\odot}$ so that the photo-$z$ errors are not excessive and so that the photo-$z$ catalogue is complete in stellar mass. Most of these objects have $25<I_{\mathrm{AB}}<28$.
We first search for any excess of galaxies around the locations of the candidate groups.  We consider cylinders with radii that are a varying multiple of the group $r_{\mathrm{rms}}$ and which have a fixed length of twice 10'000\,km/s.  We lay down 42 cylinders, one over each group, and compare the total number of massive ($>10^{10}$\,M$_{\odot}$) galaxies in these cylinders to the totals found when the 42 cylinders are laid down at positions that have the same $(z,r_{\mathrm{rms}},dv)$ but random (RA, DEC) positions, repeating these random samples 1000 times and using the variation in the random samples to give an estimate of the noise to be expected in the group sample.  

\begin{figure}[t!]
\includegraphics[scale=0.42]{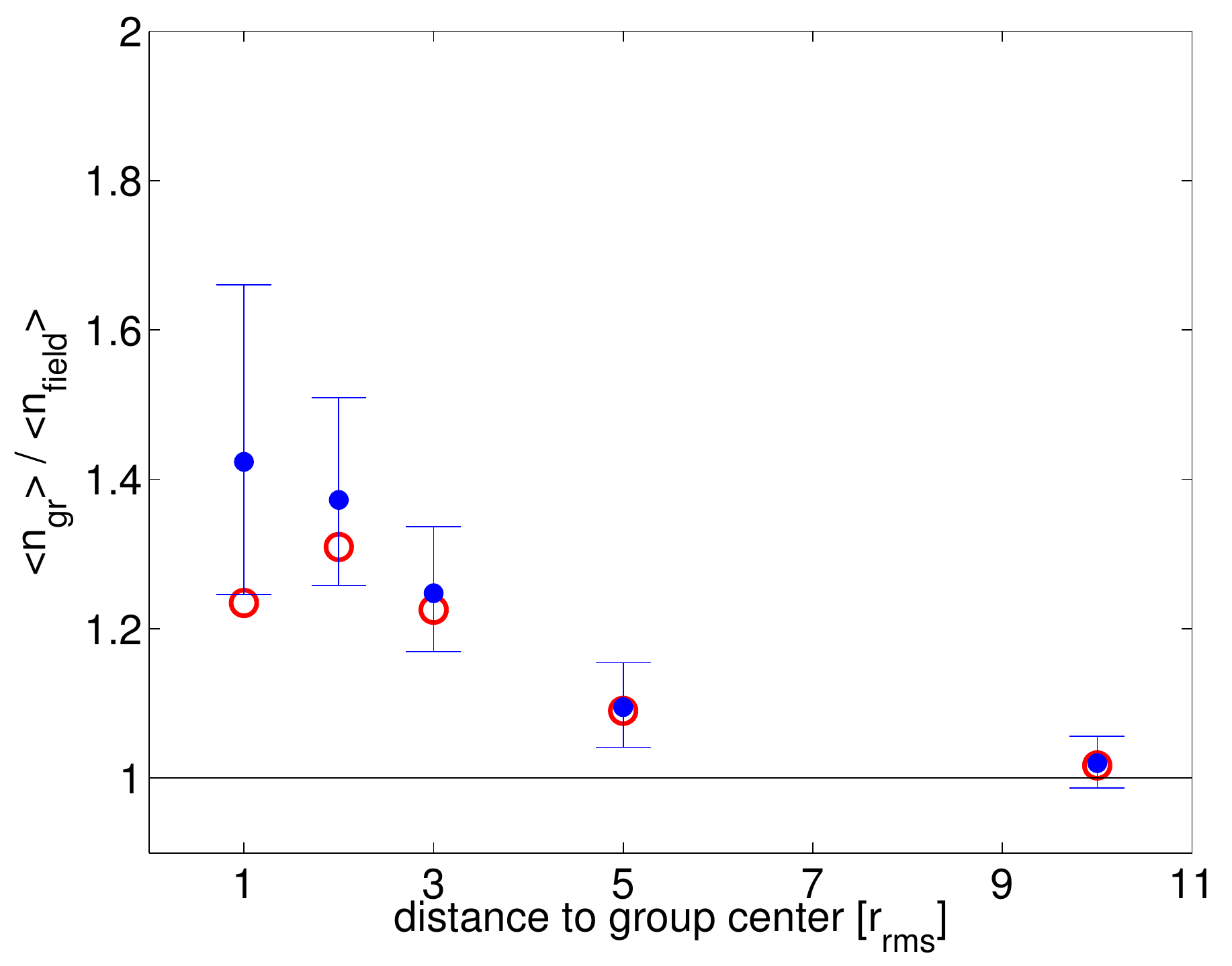}
\caption{\small The excess of high mass ($>10^{10}$\,M$_{\odot}$) galaxies from the COSMOS photo-$z$ sample around our spectroscopic candidate groups, relative to the field, as a function of the projected distance from the group in units of the $r_{\mathrm{rms}}$ of the groups as seen in cylinders of depth $\Delta v = \pm 10'000$\,km/s to accommodate photo-$z$ errors (see text for details). At the position of the candidate groups we find a projected excess of up to $\sim$40\% in the number of massive galaxies (blue filled circles). This fraction reduces to $\sim$25\% if we subtract out the already known spectroscopic members (red open circles) and also reduces to insignificance at large radii.  This concentrated mean overdensity suggests that our candidate groups indeed trace significant overdensities in the Universe.\normalsize}
\label{fig_excessHighMass}
\end{figure}

Especially at small multiples of $r_{\mathrm{rms}}$ a significant excess is seen around the candidate groups as shown in Figure \ref{fig_excessHighMass}. 
At the position of the candidate groups within a $1-2\,r_{\mathrm{rms}}$ radius we find $\sim$ 40\% more massive objects around the group positions as in the general field, whereas this fraction drops for larger radii and is consistent with unity at $\sim10\,r_{\mathrm{rms}}$, which corresponds to $\sim3\,$Mpc (physical).

This excess is only slightly reduced when the spectroscopically observed objects are excluded (red circles in Figure \ref{fig_excessHighMass}), and the excess seen in this independent dataset provides further evidence that the candidate groups catalogued in this paper are real physical associations and not just chance projections.   

\begin{figure}[t!]
\includegraphics[scale=0.42]{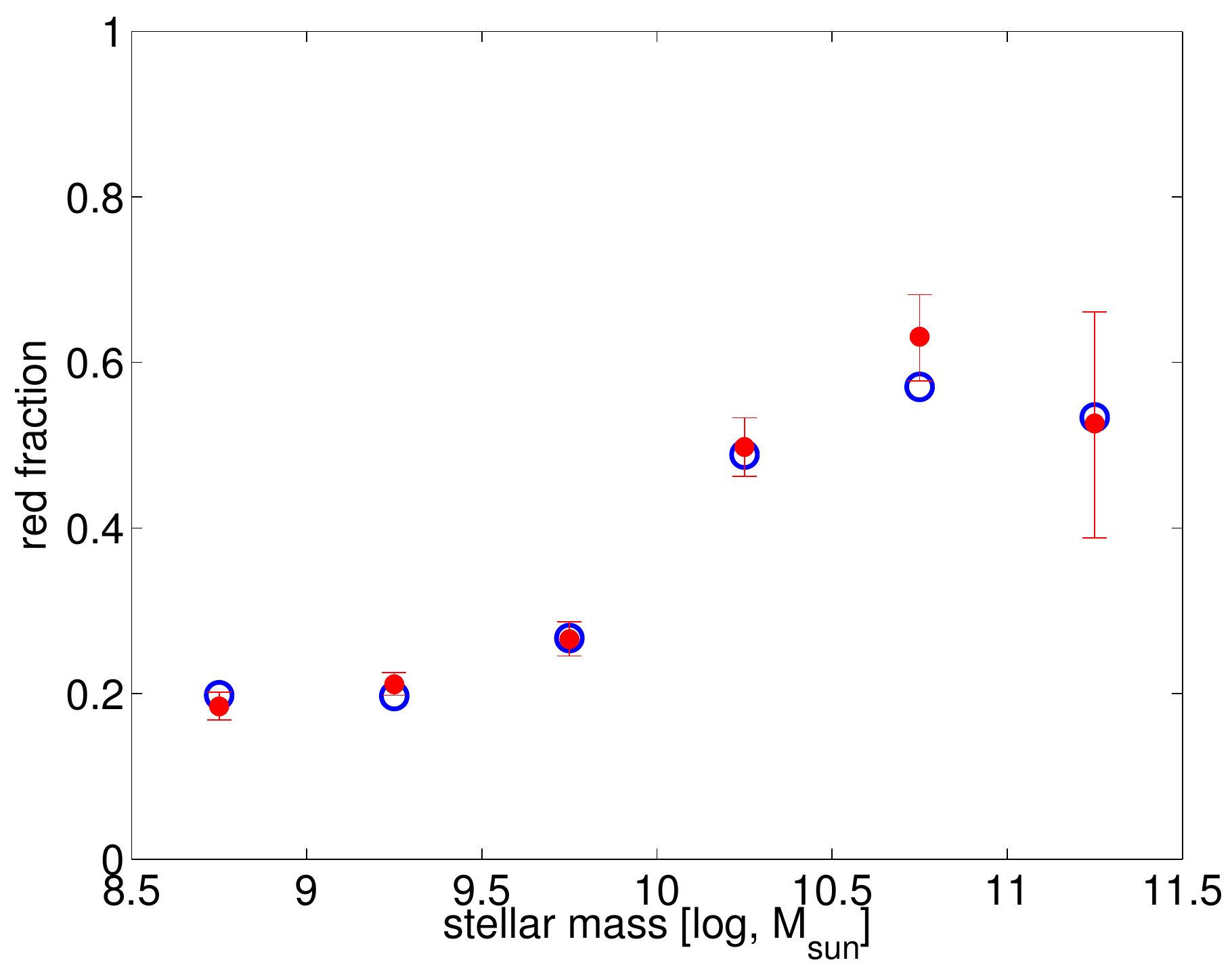}
\caption{\small The red fraction of objects in the photo-$z$ sample at the position of our candidate groups (in red) as compared to the field (in blue) as a function of stellar mass. Red galaxies are defined to have rest-frame $M_U-M_B>0.7$, using the spectral energy distributions used to estimate their photometric redshifts. We find that there is no difference in the colours for the field and the candidate groups. This is, however, not surprising if the candidate groups are only starting to assemble and if environmental differentiation is confined to satellites, as indicated at lower redshifts.\normalsize}
\label{fig_redfraction}
\end{figure}

Next we look at the distribution of colours in the photo-$z$ sample around the candidate groups with respect to the field.  For this we consider cylinders with a fixed radius of $2\,r_{\mathrm{rms}}$ and the same length of twice 10'000\,km/s as above. We define red galaxies to be galaxies with $M_U-M_B>0.7$ and consider a red fraction which is the number of red galaxies at a given stellar mass divided by the total number of galaxies at that mass. Figure \ref{fig_redfraction} 
shows the red fractions as function of stellar mass. 

It is clear that the fraction of red objects in the candidate groups and in the field, at fixed stellar mass, is essentially the same and we do not see evidence of color segregation with environment.  Of course, given the large cylinder length in redshift (of order $\pm 0.1$ around the group location), our ``group sample'' will have been heavily contaminated by unrelated foreground and background field galaxies: our overdensity of 40\% suggests that also 70\% of the photo-$z$ ``group sample'' galaxies are projected from the field.  These projected galaxies will of course heavily dilute any intrinsic color difference and we could in principle subtract these projected galaxies statistically. However, because the red fractions are so indistinguishable, we have not attempted to do this.

It is not clear that any such environmental segregation, at fixed stellar mass, should have been expected.  We have argued above that the galaxies in the candidate groups are in general unlikely, at the epoch at which we observe them, to be sharing the same dark matter halo. A correspondingly small fraction of the galaxies will be satellites, even in the larger photo-$z$ sample. 
Various papers (van den Bosch et al. 2008, Font et al. 2008,  Weinmann et al. 2009, Prescott et al. 2011, Peng et al. 2012, and many others) have presented clear evidence that all of the environmental differentiation of the galaxy population at low redshift is associated with the quenching of star-formation in satellite galaxies, and there is now also good evidence that this remains true also at redshifts approaching unity (Kovac et al. 2012 in prep.).

\section{Summary \& Conclusions}

We have applied a group-finder with linking lengths $\Delta r=500$\,kpc (physical) and $\Delta v=700$\,km/s to the zCOSMOS-deep sample of 3502 galaxies at $1.8 < z < 3.0$, yielding 42 systems with three or more members.  To try to understand what these associations likely are, and what they will probably become, we have constructed an analogous sample from 12 zCOSMOS-deep mock samples which were extracted from the Millennium simulation mock catalogues of \citet{kitzbichler07}, supplemented by a single light cone from the \citet{wang08} simulation which has a more realistic value of $\sigma_8$.  

We refer to the detected systems as ``candidate groups''.  We have introduced the following terminology in which a system in which all three detected members are in the same halo is called a ``real group'' and one in which only two are, a ``partial group''.  Candidate groups that will become real or partial groups by $z = 0$ are called ``proto-groups'' and ``partial proto-groups'' respectively.

The number of candidate groups in the simulations agrees quite well with the number in the sky.  However, analysis of the simulated candidate groups suggests that only a very small fraction, less than 0.2\% in the \citet{kitzbichler07} sample and none in the Wang et al. (2008) sample, 
already have all the detected galaxies occupying the same halo at the time of observation, i.e., are already ``real groups''.  About 8\% of the candidate groups will however already have two members within the same halo in the \citet{kitzbichler07} sample.

Furthermore, 50\% of the mock candidate groups will have assembled all three galaxies into the same halo by $z = 0$ (i.e., are ``proto-groups'' at the epoch of observation) and almost all (93\%) will have at least two galaxies in the same halo.   Only 7\% are truly random associations whose members will never occupy the same halo. The mocks suggest that the important parameter that distinguishes the fate of the candidate group is the apparent velocity dispersion $v_{\mathrm{rms}}$. For $v_{\mathrm{rms}}\lesssim300$\,km/s the fraction of system that will fully assemble all three members is above $50\%$ and for larger dispersions it is lower.  The fraction does not depend much on the projected angular size of the candidate groups.

The observed candidate groups are being seen as they begin the assembly process. Already by $z\sim1.8$ (which is the lower limit of our observational window) around $25\%$ of the candidate groups (observed at $1.8 < z < 3.0$) will be partial-groups, the bulk of them doing so within $\Delta a<0.1$ from their epoch of detection, and within $\Delta a\lesssim0.5$ most proto-groups will have evolved into real or partial groups.

If we look at today's groups and ask which of their progenitors will have been seen in our spectroscopic sample at $z > 1.8$, then we find that we should have detected $\sim 35\%$ of the progenitors of todays massive clusters (of order of $10^{14} - 10^{15}$\,M$_{\odot}$/h) already at $z\sim2$ and this would rise to $\sim 65\%$ if we had 100\% completeness in the zCOSMOS-deep spectroscopic sample.

We can roughly estimate the overdensities of the spectroscopically detected structures and find that these are substantial, consistent with the idea that these systems will soon come together into assembled systems.  

We also detect a significant overdensity in the regions of these candidate groups using independent the COSMOS photometric sample, which shows a 40\% excess in the numbers of galaxies above $10^{10}$M$_{\sun}$ at the location of our spectroscopic candidate groups as compared to the field, despite the very large sampling cylinders ($\Delta z = \pm 0.1$) required from the use of photo-$z$. We do not however detect any significant differentiation in the colours of the galaxies compared to the field.  However, we might not have expected to see such differences if most of the structures are still assembling on account of the fact that at $z < 1$ environmental differentiation of the galaxy population is confined to satellite galaxies.

\acknowledgments
\section*{Acknowledgements}
This research is based on observations undertaken at the European Southern Observatory (ESO) Very Large Telescope (VLT) under the Large Program 175.A-0839 and has been supported by the Swiss National Science Foundation (SNF).

G. Lemson is supported by Advanced Grant 246797 GALFORMOD from the European
Research Council.

The Millennium Simulation databases used in this paper and the web application providing online access to them were constructed as part of the activities of the German Astrophysical Virtual Observatory.

\clearpage

\begin{thebibliography}{99}
\bibitem[Abadi et al. (1999)]{abadi99} Abadi, M. G., et al. 1999, MNRAS, 308, 947
\bibitem[Balogh et al. (2004)]{balogh04} Balogh, M., et al. 2004, MNRAS, 348, 1355
\bibitem[Berlind et al. (2006)]{berlind06} Berlind, A. A., et al. 2006, ApJS, 167, 1
\bibitem[Blaizot et al.(2005)]{blaizot05} Blaizot, J., et al. 2005, MNRAS, 360, 159 
\bibitem[Capak et al. (2007)]{capak07} Capak, P. L., et al. 2007, ApJS, 172, 99
\bibitem[Capak et al. (2011)]{capak11} Capak, P. L., et al. 2011, Nature, 470, 233
\bibitem[Cucciati et al. (2012)]{cucciati10} Cucciati, O., et al. 2010, A\&A, 520, A42
\bibitem[Daddi et al. (2004)]{daddi04} Daddi, E., et al. 2004, ApJ, 617, 746
\bibitem[De Lucia \& Blaizot (2007)]{delucia07} De Lucia, G., \& Blaizot, J. 2007, MNRAS 375, 2
\bibitem[Dressler (1980)]{dressler80} Dressler A. 1980, ApJ, 236, 351
\bibitem[Eke et al. (2004)]{eke04} Eke, V. R., et al. 2004, MNRAS, 348, 866
\bibitem[Font et al.(2008)]{font08} Font, A. S., et al. 2008, MNRAS, 389, 1619 
\bibitem[Gerke et al. (2005)]{gerke05} Gerke, B. F., et al. 2005, ApJ, 625, 6
\bibitem[Gobat et al. (2011)]{gobat11} Gobat, R., et al. 2011, A\&A, 526, A133
\bibitem[Gunn \& Gott (1972)]{gunn72} Gunn, J. E., \& Gott, J. R. 1972, ApJ, 176, 1
\bibitem[Hopkins \& Beacom(2006)]{hopkins06} Hopkins, A. M., \& Beacom, J. F. 2006, ApJ, 651, 142 
\bibitem[Huchra \& Geller (1982)]{huchra82} Huchra, J. P., \& Geller, M. J. 1982, ApJ, 257, 423
\bibitem[Jenkins et al. (2001)]{jenkins01} Jenkins, A., et al. 2001, MNRAS, 321, 372
\bibitem[Kawata \& Mulchaey (2008)]{kawata08} Kawata, D., \& Mulchaey, J. S. 2008, ApJ, 672, L103
\bibitem[Kitzbichler \& White (2007)]{kitzbichler07} Kitzbichler , M. G., \& White, S. D. M., 2007, MNRAS, 376, 2
\bibitem[Knobel et al. (2009)]{knobel09} Knobel, C., et al. 2009, ApJ, 697, 1842
\bibitem[Knobel et al. (2012)]{knobel12} Knobel, C., et al. 2012, ApJ, 753, 121
\bibitem[Kovac et al. (2012)]{kovac12} Kovac, K., et al. 2012, in preparation
\bibitem[Larson et al. (1980)]{larson80} Larson, R. B., et al. 1980, ApJ, 237, 692
\bibitem[Lemson et al. (2006)]{lemson06_1} Lemson, G., \& the Virgo Consortium 2006, arXiv:astro-ph/0608019
\bibitem[Lemson \& Springel(2006)]{lemson06_2} Lemson, G., \& Springel, V. 2006, Astronomical Data Analysis Software and Systems XV, 351, 212
\bibitem[Lilly et al. (2007)]{lilly07} Lilly, S. J.,  et al. 2007, ApJS, 172, 7
\bibitem[Lilly et al. (2009)]{lilly09} Lilly, S. J., et al. 2009, ApJS, 184, 218
\bibitem[Lilly et al. (2012)]{lilly12} Lilly, S. J., et al. 2012, in preparation
\bibitem[Marinoni et al. (2002)]{marinoni02} Marinoni, C., et al. 2002, ApJ, 580, 122
\bibitem[Miley et al.(2006)]{miley06} Miley, G. K., et al. 2006, ApJL, 650, L29
\bibitem[Moore et al. (1995)]{moore95} Moore, B., et al. 1996, Nat, 379, 613
\bibitem[Oemler (1974)]{oemler74} Oemler, A. Jr. 1974, ApJ, 194, 1
\bibitem[Papovich et al. (2010)]{papovich10} Papovich, C., et al. 2010, ApJ, 716, 1503
\bibitem[Pasquali et al. (2010)]{pasquali10} Pasquali, A., et al. 2010, MNRAS, 407, 937
\bibitem[Peng et al. (2010)]{peng10} Peng, Y., et al. 2010, ApJ, 721, 193
\bibitem[Peng et al. (2012)]{peng12} Peng, Y., et al. 2012, ApJ, 757, 4
\bibitem[Prescott et al.(2011)]{prescott11} Prescott, M., et al. 2011, MNRAS, 417, 1374 
\bibitem[Reddy et al.(2008)]{reddy08} Reddy, N. A., et al. 2008, ApJS, 175, 48
\bibitem[Skibba (2009)]{skibba09} Skibba, R. A. 2009, MNRAS, 392, 1467
\bibitem[Spitler et al. (2012)]{spitler12} Spitler, L. R., et al. 2012, ApJL, 748, 21
\bibitem[Spitzer \& Baade (1951)]{spitzer51} Spitzer, L. Jr., \& Baade, W. 1951, ApJ, 113, 413
\bibitem[Springel et al. (2005)]{springel05} Springel, V., et al.  2005, Nature, 435, 629
\bibitem[Steidel et al. (2004)]{steidel04} Steidel, C. C., et al. 2004, ApJ, 604, 534
\bibitem[Steidel et al. (2005)]{steidel05} Steidel, C. C., et al. 2005, ApJ, 626, 44
\bibitem[Tanaka et al. (2010)]{tanaka10} Tanaka, M., et al. 2010, ApJL, 716, 152
\bibitem[Trenti et al. (2012)]{trenti12} Trenti, M., et al. 2012, ApJ, 746, 55 
\bibitem[van den Bosch et al. (2008)]{vandenBosch2008} van den Bosch, F. C., et al. 2008, MNRAS, 387, 79
\bibitem[Venemans et al.(2007)]{venemans07} Venemans, B. P., et al. 2007, A\&A, 461, 823 
\bibitem[Wang et al. (2008)]{wang08} Wang, J., et al. 2008, MNRAS, 384, 1301
\bibitem[Weinmann et al.(2009)]{weinmann09} Weinmann, S. M., et al. 2009, MNRAS, 394, 1213
\bibitem[Wolf et al.(2003)]{wolf03} Wolf, C., 2003, AAP, 408, 499
\end{thebibliography}
\end{document}